\pdfoutput=1
\documentclass[12pt,a4paper]{article}

\usepackage[T1]{fontenc}
\usepackage{lmodern}
\usepackage{lscape}
\usepackage[export]{adjustbox}
\usepackage{numprint}
\usepackage{multirow}
\usepackage{ifthen} 
\newboolean{pdflatex}
\setboolean{pdflatex}{true} 

\newboolean{articletitles}
\setboolean{articletitles}{true} 

\newboolean{uprightparticles}
\setboolean{uprightparticles}{false} 
\counterwithout{figure}{section}
\counterwithout{figure}{subsection}

\newboolean{inbibliography}
\setboolean{inbibliography}{false} 

\def\paperauthors{LHCb collaboration} 
\def\paperasciititle{B0->D K*0} 
\def\papertitle{Study of \CP violation in $\Bd \to \D \Kstar(892)^0$ decays with $\D\to \kaon\pion(\pion\pion)$, $\pion\pion(\pion\pion)$, and $\kaon\kaon$ final states } 
\def\paperkeywords{{High Energy Physics}, {LHCb}, {Gamma} , {CP Violation}} 
\def\papercopyright{\the\year\ CERN for the benefit of the LHCb collaboration} 
\def\paperlicence{CC BY 4.0 licence}
\def\paperlicenceurl{https://creativecommons.org/licenses/by/4.0/}

\usepackage[top=1in, bottom=1.25in, left=1in, right=1in]{geometry}

\usepackage{microtype}
\usepackage{lineno}  
\usepackage{xspace} 
\usepackage{caption} 

\usepackage{graphicx}  
\usepackage{color}
\usepackage{colortbl}
\graphicspath{{./figs/}} 

\usepackage{amsmath} 
\usepackage{amssymb}
\usepackage{amsfonts}
\usepackage{upgreek} 

\newcommand*\patchAmsMathEnvironmentForLineno[1]{%
\expandafter\let\csname old#1\expandafter\endcsname\csname #1\endcsname
\expandafter\let\csname oldend#1\expandafter\endcsname\csname
end#1\endcsname
 \renewenvironment{#1}%
   {\linenomath\csname old#1\endcsname}%
   {\csname oldend#1\endcsname\endlinenomath}%
}
\newcommand*\patchBothAmsMathEnvironmentsForLineno[1]{%
  \patchAmsMathEnvironmentForLineno{#1}%
  \patchAmsMathEnvironmentForLineno{#1*}%
}
\AtBeginDocument{%
\patchBothAmsMathEnvironmentsForLineno{equation}%
\patchBothAmsMathEnvironmentsForLineno{align}%
\patchBothAmsMathEnvironmentsForLineno{flalign}%
\patchBothAmsMathEnvironmentsForLineno{alignat}%
\patchBothAmsMathEnvironmentsForLineno{gather}%
\patchBothAmsMathEnvironmentsForLineno{multline}%
\patchBothAmsMathEnvironmentsForLineno{eqnarray}%
}

\usepackage{hyperxmp}

\usepackage[pdftex,
            pdfauthor={\paperauthors},
            pdftitle={\paperasciititle},
            pdfkeywords={\paperkeywords},
            pdfcopyright={Copyright (C) \papercopyright},
            pdflicenseurl={\paperlicenceurl}]{hyperref}

\usepackage[colorinlistoftodos,textsize=scriptsize]{todonotes}

\usepackage[bottom,flushmargin,hang,multiple]{footmisc}

\usepackage[all]{hypcap} 

\usepackage{xspace} 
\usepackage{upgreek}

\def\MagUp {\mbox{\em Mag\kern -0.05em Up}\xspace}

\ifthenelse{\boolean{uprightparticles}}%
{

 \def\Ppi         {\ensuremath{\uppi}\xspace}                 
                  
 \def\Prho        {\ensuremath{\uprho}\xspace}

 \def\PDelta      {\ensuremath{\Delta}\xspace}                 
 \def\PXi         {\ensuremath{\Xi}\xspace}                 
 \def\PLambda     {\ensuremath{\Lambda}\xspace}                 
 \def\PSigma      {\ensuremath{\Sigma}\xspace}                 
 \def\POmega      {\ensuremath{\Omega}\xspace}                 
 \def\PUpsilon    {\ensuremath{\Upsilon}\xspace}
 \let\oldPi\Pi
 \def\PPi         {\ensuremath{\oldPi}\xspace}

 \def\PB      {\ensuremath{\mathrm{B}}\xspace}                 
                  
 \def\PD      {\ensuremath{\mathrm{D}}\xspace}

 \def\PK      {\ensuremath{\mathrm{K}}\xspace}

 \def\PW      {\ensuremath{\mathrm{W}}\xspace}

 \def\Pb      {\ensuremath{\mathrm{b}}\xspace}                 
 \def\Pc      {\ensuremath{\mathrm{c}}\xspace}

 \def\Pi      {\ensuremath{\mathrm{i}}\xspace}

 \def\Pp      {\ensuremath{\mathrm{p}}\xspace}

 \def\Ps      {\ensuremath{\mathrm{s}}\xspace}                 
                  
 \def\Pu      {\ensuremath{\mathrm{u}}\xspace}

 \def\thebaroffset{0.0em}
}
{

 \def\Ppi         {\ensuremath{\pi}\xspace}                 
                  
 \def\Prho        {\ensuremath{\rho}\xspace}

 \mathchardef\PDelta="7101
 \mathchardef\PXi="7104
 \mathchardef\PLambda="7103
 \mathchardef\PSigma="7106
 \mathchardef\POmega="710A
 \mathchardef\PUpsilon="7107
 \mathchardef\PPi="7105
                  
 \def\PB      {\ensuremath{B}\xspace}                 
                  
 \def\PD      {\ensuremath{D}\xspace}

 \def\PK      {\ensuremath{K}\xspace}

 \def\PW      {\ensuremath{W}\xspace}

 \def\Pb      {\ensuremath{b}\xspace}                 
 \def\Pc      {\ensuremath{c}\xspace}

 \def\Pi      {\ensuremath{i}\xspace}

 \def\Pp      {\ensuremath{p}\xspace}

 \def\Ps      {\ensuremath{s}\xspace}                 
                  
 \def\Pu      {\ensuremath{u}\xspace}

 \def\thebaroffset{0.18em}
}
\newcommand{\offsetoverline}[2][\thebaroffset]{\kern #1\overline{\kern -#1 #2}}%

\makeatletter
\ifcase \@ptsize \relax
  \newcommand{\miniscule}{\@setfontsize\miniscule{4}{5}}
\or
  \newcommand{\miniscule}{\@setfontsize\miniscule{5}{6}}
\or
  \newcommand{\miniscule}{\@setfontsize\miniscule{5}{6}}
\fi
\makeatother

\DeclareRobustCommand{\optbar}[1]{\shortstack{{\miniscule (\rule[.5ex]{1.25em}{.18mm})}
  \\ [-.7ex] $#1$}}

\def\Wpm    {{\ensuremath{\PW^\pm}}\xspace}

\def\uquark    {{\ensuremath{\Pu}}\xspace}
\def\uquarkbar {{\ensuremath{\overline \uquark}}\xspace}

\def\squark    {{\ensuremath{\Ps}}\xspace}

\def\cquark    {{\ensuremath{\Pc}}\xspace}
\def\cquarkbar {{\ensuremath{\overline \cquark}}\xspace}

\def\bquark    {{\ensuremath{\Pb}}\xspace}

\def\pion   {{\ensuremath{\Ppi}}\xspace}
\def\piz    {{\ensuremath{\pion^0}}\xspace}
\def\pip    {{\ensuremath{\pion^+}}\xspace}
\def\pim    {{\ensuremath{\pion^-}}\xspace}
\def\pipm   {{\ensuremath{\pion^\pm}}\xspace}

\def\rhomeson {{\ensuremath{\Prho}}\xspace}
\def\rhoz     {{\ensuremath{\rhomeson^0}}\xspace}

\def\kaon    {{\ensuremath{\PK}}\xspace}
\def\Kbar    {{\ensuremath{\offsetoverline{\PK}}}\xspace}

\def\KorKbar {\kern \thebaroffset\optbar{\kern -\thebaroffset \PK}{}\xspace}

\def\Kp      {{\ensuremath{\kaon^+}}\xspace}
\def\Km      {{\ensuremath{\kaon^-}}\xspace}
\def\Kpm     {{\ensuremath{\kaon^\pm}}\xspace}
\def\Kmp     {{\ensuremath{\kaon^\mp}}\xspace}
\def\KS      {{\ensuremath{\kaon^0_{\mathrm{S}}}}\xspace}

\def\Kstarz  {{\ensuremath{\kaon^{*0}}}\xspace}
\def\Kstarzb {{\ensuremath{\Kbar{}^{*0}}}\xspace}
\def\Kstar   {{\ensuremath{\kaon^*}}\xspace}
\def\Kstarb  {{\ensuremath{\Kbar{}^*}}\xspace}

\def\Dbar    {{\ensuremath{\offsetoverline{\PD}}}\xspace}
\def\D       {{\ensuremath{\PD}}\xspace}

\def\DorDbar {\kern \thebaroffset\optbar{\kern -\thebaroffset \PD}\xspace}
\def\Dz      {{\ensuremath{\D^0}}\xspace}
\def\Dzb     {{\ensuremath{\Dbar{}^0}}\xspace}
\def\Dp      {{\ensuremath{\D^+}}\xspace}
\def\Dm      {{\ensuremath{\D^-}}\xspace}

\def\DpDm    {\ensuremath{\Dp {\kern -0.16em \Dm}}\xspace}
\def\Dstar   {{\ensuremath{\D^*}}\xspace}

\def\B       {{\ensuremath{\PB}}\xspace}
\def\Bbar    {{\ensuremath{\offsetoverline{\PB}}}\xspace}

\def\BorBbar {\kern \thebaroffset\optbar{\kern -\thebaroffset \PB}\xspace}
\def\Bz      {{\ensuremath{\B^0}}\xspace}
\def\Bzb     {{\ensuremath{\Bbar{}^0}}\xspace}
\def\Bd      {{\ensuremath{\B^0}}\xspace}
\def\Bdb     {{\ensuremath{\Bbar{}^0}}\xspace}
\def\BdorBdbar {\kern \thebaroffset\optbar{\kern -\thebaroffset \Bd}\xspace}
\def\Bu      {{\ensuremath{\B^+}}\xspace}

\def\Bpm     {{\ensuremath{\B^\pm}}\xspace}

\def\Bs      {{\ensuremath{\B^0_\squark}}\xspace}

\def\BsorBsbar {\kern \thebaroffset\optbar{\kern -\thebaroffset \Bs}\xspace}

\def\Bds     {{\ensuremath{\B_{(\squark)}^0}}\xspace}

\def\BdorBs  {\Bds}

\def\Y#1S{\ensuremath{\PUpsilon{(#1S)}}\xspace}

\def\proton      {{\ensuremath{\Pp}}\xspace}

\def\LorLbar     {\kern \thebaroffset\optbar{\kern -\thebaroffset \PLambda}\xspace}

\def\to                 {\ensuremath{\rightarrow}\xspace}

\def\CP                {{\ensuremath{C\!P}}\xspace}

\def\AT#1     {\ensuremath{A_{\mathrm{T}}^{#1}}\xspace}           

\def\C#1      {\ensuremath{\mathcal{C}_{#1}}\xspace}                       
\def\Cp#1     {\ensuremath{\mathcal{C}_{#1}^{'}}\xspace}                    
\def\Ceff#1   {\ensuremath{\mathcal{C}_{#1}^{\mathrm{(eff)}}}\xspace}        
\def\Cpeff#1  {\ensuremath{\mathcal{C}_{#1}^{'\mathrm{(eff)}}}\xspace}       
\def\Ope#1    {\ensuremath{\mathcal{O}_{#1}}\xspace}                       
\def\Opep#1   {\ensuremath{\mathcal{O}_{#1}^{'}}\xspace}                    

\newcommand{\nospaceunit}[1]{\ensuremath{\text{#1}}}       
\newcommand{\aunit}[1]{\ensuremath{\text{\,#1}}}       

\newcommand{\tev}{\aunit{Te\kern -0.1em V}\xspace}
\newcommand{\gev}{\aunit{Ge\kern -0.1em V}\xspace}
\newcommand{\mev}{\aunit{Me\kern -0.1em V}\xspace}
\newcommand{\kev}{\aunit{ke\kern -0.1em V}\xspace}
\newcommand{\ev}{\aunit{e\kern -0.1em V}\xspace}
 
\newcommand{\mevc}{\ensuremath{\aunit{Me\kern -0.1em V\!/}c}\xspace}
\newcommand{\gevc}{\ensuremath{\aunit{Ge\kern -0.1em V\!/}c}\xspace}
\newcommand{\mevcc}{\ensuremath{\aunit{Me\kern -0.1em V\!/}c^2}\xspace}
\newcommand{\gevcc}{\ensuremath{\aunit{Ge\kern -0.1em V\!/}c^2}\xspace}

\def\mum  {\ensuremath{\,\upmu\nospaceunit{m}}\xspace}

\def\gsim{{~\raise.15em\hbox{$>$}\kern-.85em
          \lower.35em\hbox{$\sim$}~}\xspace}
\def\lsim{{~\raise.15em\hbox{$<$}\kern-.85em
          \lower.35em\hbox{$\sim$}~}\xspace}

\def\evtgen     {\mbox{\textsc{EvtGen}}\xspace}

\def\geant      {\mbox{\textsc{Geant4}}\xspace}

\def\photos     {\mbox{\textsc{Photos}}\xspace}

\def\pythia     {\mbox{\textsc{Pythia}}\xspace}

\def\tell1  {TELL1\xspace}
\def\ukl1   {UKL1\xspace}

\newcommand{\eg}{\mbox{\itshape e.g.}\xspace}

\newcommand{\etc}{\mbox{\itshape etc.}\xspace}

\newcommand{\lhcborcid}[1]{\href{https://orcid.org/#1}{\hspace*{0.1em}\raisebox{-0.45ex}{\includegraphics[width=1em]{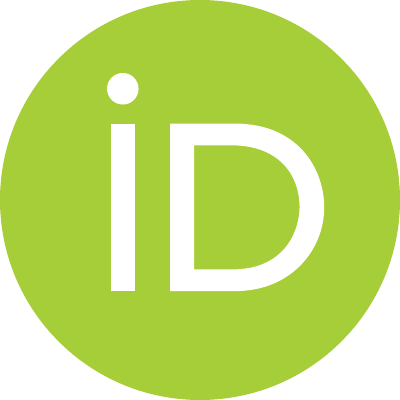}}}}

\usepackage{cite} 
\usepackage{mciteplus}

\usepackage{longtable} 

\PassOptionsToPackage{unicode}{hyperref}
\PassOptionsToPackage{naturalnames}{hyperref}
\PassOptionsToPackage{amsmath}{hyperref}
\begin{document}

\renewcommand{\thefootnote}{\fnsymbol{footnote}}
\setcounter{footnote}{1}


\begin{titlepage}
\pagenumbering{roman}

\vspace*{-1.5cm}
\centerline{\large EUROPEAN ORGANIZATION FOR NUCLEAR RESEARCH (CERN)}
\vspace*{1.5cm}
\noindent
\begin{tabular*}{\linewidth}{lc@{\extracolsep{\fill}}r@{\extracolsep{0pt}}}
\ifthenelse{\boolean{pdflatex}}
{\vspace*{-1.5cm}\mbox{\!\!\!\includegraphics[width=.14\textwidth]{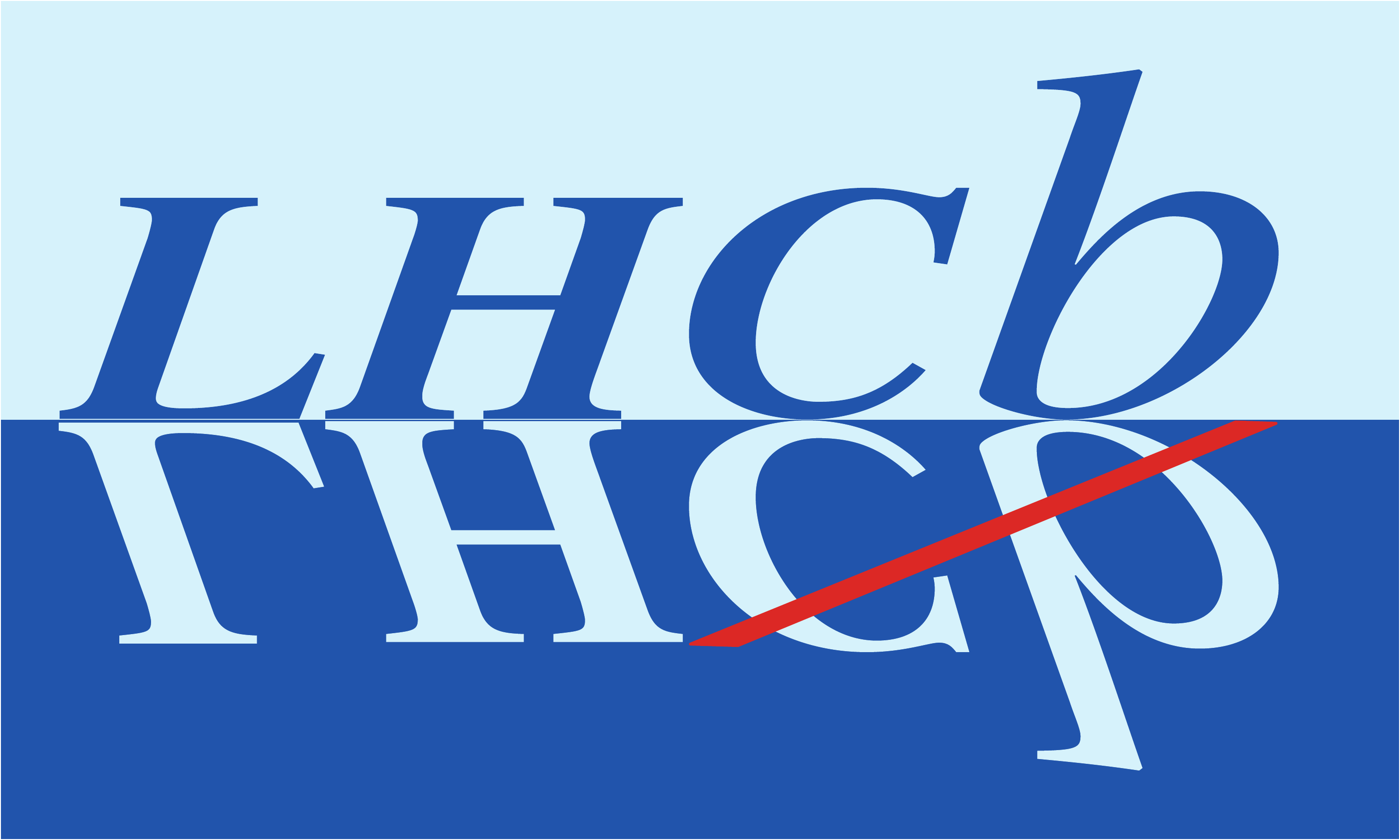}} & &}%
{\vspace*{-1.2cm}\mbox{\!\!\!\includegraphics[width=.12\textwidth]{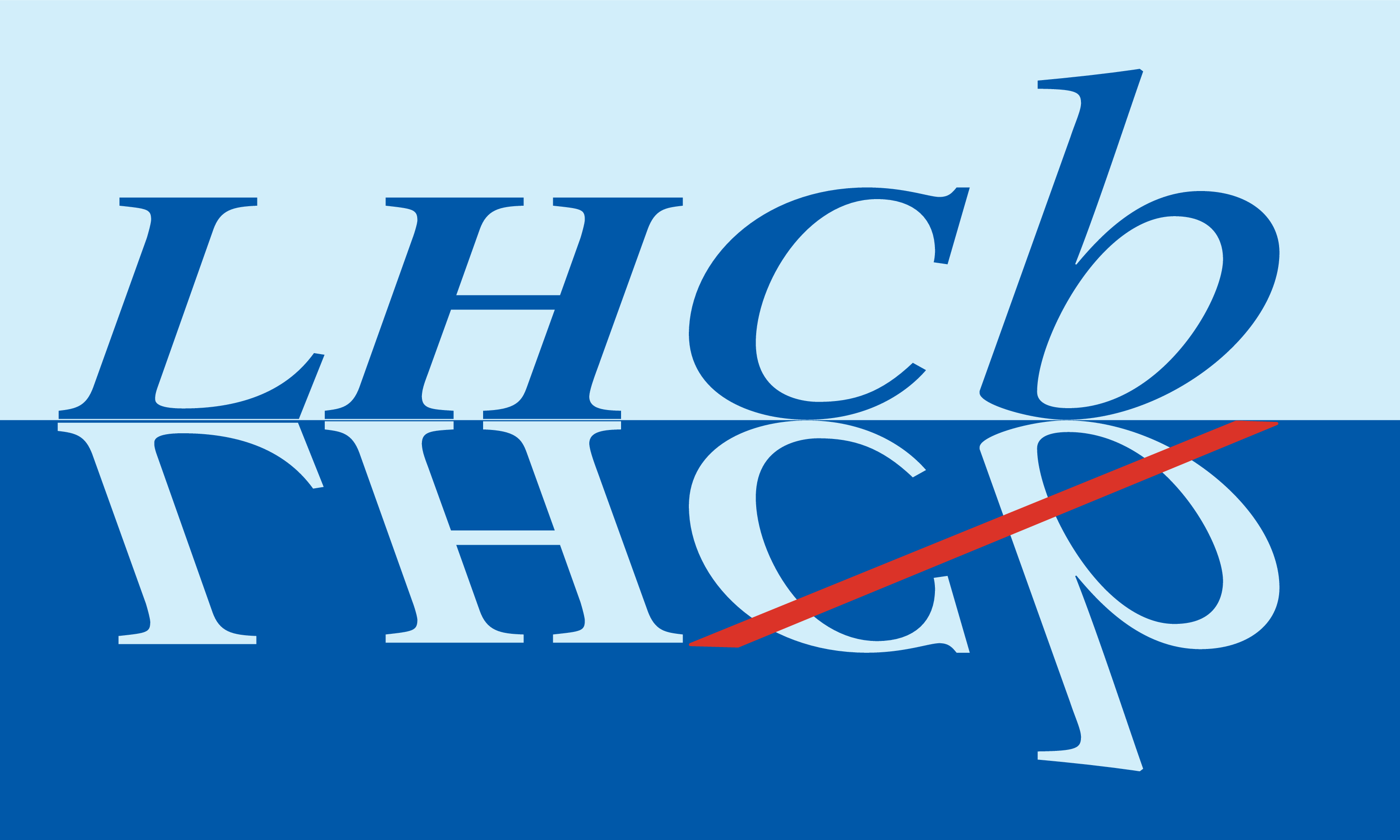}} & &}%
\\
 & & CERN-EP-2024-007 \\  
 & & LHCb-PAPER-2023-040 \\  
 & & \today\\ 
 & & \\
\end{tabular*}

\vspace*{4.0cm}

{\normalfont\bfseries\boldmath\huge
\begin{center}
  \papertitle 
\end{center}
}

\vspace*{2.0cm}

\begin{center}
\paperauthors\footnote{Authors are listed at the end of this paper.}
\end{center}

\vspace{\fill}

\begin{abstract}
  \noindent
A measurement of \CP-violating observables associated with the interference of $\Bd\to \Dz \Kstar(892)^0$ and $\Bd\to \Dzb \Kstar(892)^0$ decay amplitudes is performed in the $\Dz \to \Kmp\pipm(\pip\pim),$ $\Dz \to \pip\pim(\pip\pim)$, and $\Dz\to \Kp\Km$ final states using data collected by the LHCb experiment corresponding to an integrated luminosity of $9$~$\text{fb}^{-1}$. \CP-violating observables related to the interference of $\Bs\to \Dz \Kstarb(892)^0$ and $\Bs\to \Dzb \Kstarb(892)^0$ are also measured, but no evidence for interference is found. The \Bd observables are used to constrain the parameter space of the CKM angle $\gamma$ and the hadronic parameters $r_{\Bd}^{\D\Kstar}$ and $\delta_{\Bd}^{\D\Kstar}$ with inputs from other measurements. In a combined analysis, these measurements allow for four solutions in the parameter space, only one of which is consistent with the world average.
\end{abstract}

\vspace*{2.0cm}

\begin{center}
  Published in JHEP 05(2024) 025
\end{center}

\vspace{\fill}

{\footnotesize 
\centerline{\copyright~\papercopyright. \href{\paperlicenceurl}{\paperlicence}.}}
\vspace*{2mm}

\end{titlepage}


\newpage
\setcounter{page}{2}
\mbox{~}
%
%
%
%


\renewcommand{\thefootnote}{\arabic{footnote}}
\setcounter{footnote}{0}



\pagestyle{plain} 
\setcounter{page}{1}
\pagenumbering{arabic}


\section{Introduction}
\label{sec:Introduction}

The only observed phenomena of \CP violation to date are attributed to complex phases in the Cabibbo-Kobayashi-Maskawa (CKM) matrix elements\cite{Cabibbo:1963yz,Kobayashi:1973fv}, which describe the interactions mediated by $\Wpm$ bosons between quarks of different flavour and form a three-by-three unitary matrix. The unitarity of the CKM matrix results in a set of constraints between the matrix elements. One such relation, particularly relevant to decays of $\Bd$ mesons, is commonly parameterised in terms of three angles $\alpha$, $\beta$, and $\gamma$ which sum to $180^\circ$. The angle $\gamma$ can be measured from the interference of $\bquark\to \cquark\uquarkbar \squark$ and $\bquark\to \cquarkbar \uquark \squark$ amplitudes with negligible theoretical uncertainties\cite{Brod:2013sga}. Precise measurements of $\gamma$ provide stringent tests of the Standard Model's requirement of CKM unitarity and thus its description of \CP violation.

 The determination of $\gamma$ through direct measurements of $\bquark\to \cquark\uquarkbar \squark$ and $\bquark\to \cquarkbar \uquark \squark$ interference is currently driven by the LHCb experiment. The world average of direct measurements from the HFLAV collaboration is $\gamma=\left(66.2^{+3.4}_{-3.6}\right)^\circ$\cite{HeavyFlavorAveragingGroup:2022wzx}, and is dominated by $\Bu\to \D\Kp$ decays that have been analysed with the full proton-proton ($pp$) collision datasets collected by the LHCb detector in 2011--2012 (Run 1) and 2015--2018 (Run 2)~\cite{LHCb-CONF-2022-003}.  Here and throughout the article, $\D$ is used to represent a superposition of $\Dz$ and $\Dzb$ decays, with a similar convention for $\Dstar$. The value of $\gamma$ can be determined indirectly by assuming CKM unitarity and performing a global fit to all measurements relating to the CKM matrix. With such techniques, the UTFit collaboration determined $\gamma=\left(64.9\pm1.4\right)^\circ$\cite{UTfit:2022hsi} and the CKMFitter collaboration determined $\gamma=\left(65.5^{+1.3}_{-1.2}\right)^\circ$\cite{CKMfitter2005}. While the results of the direct and indirect determinations of $\gamma$ are in agreement, the comparison is limited by the precision of direct measurements. As such, additional direct measurements of $\gamma$ are required to provide more stringent tests of CKM unitarity. Despite the smaller branching fraction of $\Bd\to \D \Kstar(892)^0$ decays compared to $\Bu\to \D \Kp$ decays, a competitive precision on $\gamma$ can be achieved due to the larger interference effects in these decays compared to $\Bu$ decays\cite{Gershon_2009}. The \CP violation in these decays only depends on the flavour at decay, and thus can be studied independent of decay time.

This article presents the measurement of \CP-violating observables in the decays $\Bd\to \D \Kstar(892)^0$ and $\Bdb\to \D \Kstarb(892)^0$ where the \D meson is reconstructed through the $ \Kmp\pipm,\;\Kp\Km,\;\pip\pim,\;\Kmp\pipm\pip\pim, \text{ and } \pip\pim\pip\pim$ decay modes with Run 1 and Run~2 proton-proton collision data corresponding to an integrated luminosity of $9$~$\text{fb}^{-1}$ collected at centre-of-mass energies between $7$ and $13\tev$. The $\Kstar(892)^0$ meson is referred to throughout as $\Kstarz$ and is implied to decay to a $\Kp\pim$ final state. The charge of the kaon child from the $\Kstarz$ candidate is used to determine the flavour of the parent $\B$ meson. The interpretation of the measured \CP-violating observables in terms of $\gamma$ and hadronic parameters for $\Bd\to D \Kstarz$ decays is also presented. This work examines the so-called Atwood-Dunietz-Soni (ADS) final states\cite{Atwood:2000ck}, which experience interference in the admixture of Cabibbo-favoured $\Dz\to \Km\pip(\pip\pim)$ and doubly Cabibbo-suppressed $\Dzb\to \Km\pip(\pip\pim)$ decays. The notation $\kaon\pion(\pion\pion)$ is used to refer to the final state where the kaon child of the $\D$ candidate  has the same charge as the kaon child of the $\Kstarz$ candidate, while $\pion \kaon(\pion\pion)$ refers to the final state where the two kaons have opposite charge. The \CP-eigenstate final states $\D\to \pip\pim$ and $\D\to \Kp \Km$ are also studied and referred to as Gronau-London-Wilder (GLW) decay modes \cite{Gronau:1991dp,Gronau:1990ra}. An extension to the GLW method \cite{FPlusPiPiPi0} allows the inclusion of the $\D \to \pip \pim \pip \pim$ decay mode which is predominantly \CP-even. 

These two classes of final states provide complementary sensitivity to the fundamental parameters of interest: the weak phase, $\gamma$, the ratio of amplitudes between the \linebreak$\Bd\to \Dzb \Kstarz$ and $\Bd\to \Dz\Kstarz$ decays, $r_{\Bd}^{\D\Kstar}$, and the strong phase difference between the two amplitudes
, $\delta_{\Bd}^{\D\Kstar}$. The interference can introduce asymmetries in the rates between $\Bdb\to \D \Kstarzb$ and $\Bd\to \D\Kstarz$ decays, and modulate the charge-integrated decay rates of $\Bd\to \D\Kstarz$ for each $\D$ meson decay. The effects of interference in the $\Bd\to D[\kaon\pion(\pion\pion)]\Kstarz$ final state are expected to be small compared to the predicted experimental sensitivity, and so this final state is used as a normalisation channel to probe interference effects in the other final states. This strategy benefits from the cancellation of a large number of systematic uncertainties related to the reconstruction and selection of signal candidates. For the $\Bd\to \D[\pion \kaon (\pion\pion)]\Kstarz$ final state, the parameters
\begin{equation}
	\mathcal R^+_{\pion\kaon(\pion\pion)}\equiv\frac{\Gamma\left(\Bd\to \D\left[\pion\kaon(\pion\pion)\right]\Kstarz\right)}{\Gamma\left(\Bd\to \D\left[\kaon\pion(\pion\pion)\right]\Kstarz\right)}\quad \text{ and }\quad\mathcal R^-_{\pion\kaon(\pion\pion)}\equiv\frac{\Gamma\left(\Bdb\to \D\left[\pion\kaon(\pion\pion)\right]\Kstarzb\right)}{\Gamma\left(\Bdb\to \D\left[\kaon\pion(\pion\pion)\right]\Kstarzb\right)}
\end{equation}
are measured. The asymmetry of the $\Bd\to \D[ \kaon\pion(\pion\pion)]\Kstarz$ final state,
\begin{equation}
	\mathcal A_{\kaon\pion}\equiv \frac{\Gamma\big(\Bdb \to 
 \D[\kaon\pion(\pion\pion)] \Kstarzb\big)-\Gamma\big(\Bd \to \D[\kaon\pion(\pion\pion)] \Kstarz\big)}{\Gamma\big(\Bdb \to \D[\kaon\pion(\pion\pion)] \Kstarzb\big)+\Gamma\big(\Bd \to \D[\kaon\pion(\pion\pion)] \Kstarz\big)} 
	\label{eq:AKpi},
\end{equation}
is also measured.

For the GLW final states, the flavour-integrated decay rates relative to the \mbox{$\Bd\to \D[\kaon\pion(\pion\pion)]\Kstarz$} decay are measured, leading to the cancellation of most systematic uncertainties related to the estimation of selection efficiencies. The asymmetries do not require a normalisation channel for the cancellation of systematic uncertainties, so the observables of interest are 

\begin{equation}
\small
	\mathcal{R}^{hh(\pion\pion)}_{CP}\equiv \frac{\Gamma\big(\Bdb \to \D[hh(\pion\pion)]\Kstarzb\big)+\Gamma\big(\Bd \to \D[hh(\pion\pion)]\Kstarz\big)}{\Gamma\big(\Bdb \to \D[\kaon\pion(\pion\pion)]\Kstarzb\big)+\Gamma\big(\Bd \to \D[\kaon\pion(\pi\pi)]\Kstarz\big)}\times \frac{\mathcal B \big(\Dz\to \kaon\pion(\pion\pion)\big)}{\mathcal B \big(\Dz\to hh(\pion\pion)\big)}
	\label{eq:Rhh}
\end{equation}
and
\begin{equation}
	\mathcal A_{CP}^{hh(\pion\pion)}\equiv \frac{\Gamma\big(\Bdb \to \D[hh(\pion\pion)] \Kstarzb\big)-\Gamma\big(\Bd \to \D[hh(\pion\pion)] \Kstarz\big)}{\Gamma\big(\Bdb \to \D[hh(\pion\pion)] \Kstarzb\big)+\Gamma\big(\Bd \to \D[hh(\pion\pion)] \Kstarz\big)} 
	\label{eq:Acp},
\end{equation}
where $hh(\pion\pion)$ refers to $\kaon\kaon$, $\pion\pion$, or $\pion\pion\pion\pion$.

The \CP-violating observables in the decay $\Bs\to \D\Kstarzb$  are also measured in this analysis, and can be parameterised in an identical fashion.
However, the effect of the interference in this process is expected to be negligible in all of the observed decay modes, as the parameter directly proportional to the interference effects, $r_{\Bs}^{\D\Kstar}$, is estimated to be roughly an order of magnitude smaller than the corresponding $\Bd$ parameter based on the known CKM elements \cite{HeavyFlavorAveragingGroup:2022wzx}. Similar notation is used to refer to the $\Bs$ observables throughout the article, with an additional $s$ in the subscript.

All of these observables have been measured previously by the LHCb collaboration using the data collected in Run 1 and 2015--2016, except the ones associated with the $\D\to \pion\pion\pion\pion$ decay channel, for which only the 2015--2016 data was analysed \cite{LHCb-PAPER-2019-021}. The measurements presented in this article supersede the results of the previous analyses.

The rest of the article is organised as follows. The LHCb detector and simulation is described in Sect.~\ref{sec:Detector}. The selection requirements placed on signal candidates are discussed in Sect.~\ref{sec:Selection}. The determination of \CP-violating observables from the signal candidates passing selection requirements is discussed in Sect.~\ref{sec:Fit}. The interpretation of the determined \CP-violating observables in terms of the fundamental physics parameters and concluding remarks are presented in Sect.~\ref{sec:Interpretation}.

\section{LHCb detector and simulation}
\label{sec:Detector}

The LHCb detector \cite{LHCb-DP-2014-002,LHCb-DP-2008-001} is a single-arm forward spectrometer covering the pseudorapidity range $2 < \eta < 5$, designed for the study of particles containing $\bquark$ or $\cquark$ quarks.
The detector includes a high-precision tracking system consisting of a silicon-strip vertex detector surrounding the $\proton\proton$ interaction region \cite{LHCb-DP-2014-001}, a large-area silicon-strip detector located upstream of a dipole magnet with a bending power of about $4\,\textrm{Tm}$, and three
stations of silicon-strip detectors and straw drift tubes \cite{LHCb-DP-2013-003} placed downstream of the
magnet. The tracking system provides a measurement of the momentum, $p$, of charged
particles with a relative uncertainty that varies from $0.5\%$ at low momentum to $1.0\%$ at
$200\gevc$. The minimum distance of a track to a primary $\proton\proton$ collision vertex (PV), the
impact parameter (IP), is measured with a resolution of $(15 + 29/p_\text{T})\mum$, where $p_\text{T}$ is the
component of the momentum transverse to the beam, in $\gevc$. Different types of charged hadrons are distinguished using information from two ring-imaging Cherenkov (RICH)
detectors \cite{LHCb-DP-2012-003,LHCb-DP-2018-001}. Photons, electrons and hadrons are identified by a calorimeter system consisting of scintillating-pad and preshower detectors, an electromagnetic and a hadronic calorimeter. Muons are identified by a system composed of alternating layers of iron and multiwire proportional chambers \cite{LHCb-DP-2012-002}. 

The events considered in the analysis are triggered at the hardware level either when one of the final-state tracks of the signal decay deposits enough energy in the calorimeter system, or when one of the other particles in the event, not reconstructed as part of the signal candidate, fulfils any trigger requirement. At the software level, it is required
that at least one particle has high $p_\text{T}$ and high $\chi^{2}_\text{IP}$, where $\chi^{2}_\text{IP}$ is defined as the difference in the PV fit $\chi^{2}$ with and without the inclusion of that particle. A multivariate algorithm \cite{BBDT} is used to identify displaced vertices consistent with being a two-, three-, or
four-track $\bquark$-hadron decay. The PVs are fitted with and without the $\B$ candidate tracks, and the PV that gives the smallest $\chi^{2}_\text{IP}$ is associated with the $\B$ candidate.

Simulation is used to model the reconstructed mass distributions of the signal and background contributions and determine their selection efficiencies. In the simulation,
proton-proton collisions are generated using \pythia \cite{Sjostrand:2006za} with a specific LHCb configuration \cite{LHCb-PROC-2010-056}.
Decays of unstable particles are described by \evtgen \cite{Lange:2001uf}, in which final-state radiation is
generated using \photos \cite{davidson2015photos}. The simulation of interactions of the generated particles with the detector
and its response is implemented using the \geant toolkit \cite{Agostinelli:2002hh,Allison:2006ve} as described in Ref.~\cite{LHCb-PROC-2011-006}.
Some subdominant sources of background are generated with RapidSim, a fast simulation \cite{Cowan:2016tnm} that mimics the geometric acceptance and tracking resolution of the LHCb detector as well as the dynamics of the decay via \evtgen.

\section{Selection, efficiencies, and production and detection asymmetries}
\label{sec:Selection}
Signal $\B$-meson candidates are built from a $\Kstarz\to \Kp\pim$ candidate and a $\D$-meson candidate. Selection requirements identical to those from Ref.~\cite{LHCb-PAPER-2015-059} are placed on the $\Kstarz\to \Kp\pim$ candidate to select the $\Kstarz$ resonance: the reconstructed mass of the $\Kstarz$ candidate  must be within $50 \mevcc$ of the known $\Kstarz$ meson mass \cite{PDG2022} and a requirement is made on the angle $\theta^*$ between the $\Kstarz$ kaon child momentum and the $\Bd$ candidate in the $\Kstarz$ rest frame of $|\cos \theta^*|>0.4$. The reconstructed mass of the $\D$ candidate is required to be within $25 \mevcc$ of the known $\Dz$ meson mass\cite{PDG2022}.

With the above selections in place, boosted decision tree classifiers with gradient boost (BDT) \cite{Therhaag:2010zz} are trained to discriminate between signal candidates and combinatorial background. Two BDT classifiers are trained for each final state, one for each run period. The same BDT classifier is shared for each run period between the $\Bd\to [\kaon\pion(\pion\pion)]\Kstarz$ and $\Bd\to [\pion \kaon(\pion\pion)]\Kstarz$ final states. This gives ten separate BDT classifiers in total, each trained using simulated signal samples and background samples from data with a $\Bd$ candidate reconstructed mass within the range 5800--5960\mevcc. The classifiers are trained using the angle between the direction of the reconstructed \B momentum and the direction
defined by the primary and secondary vertices, the $\chi^{2}_\text{IP}$ of the $\Bd$ and $\D$ candidates, the reduced $\chi^2$ of the $\Bd$ vertex fit, the $\chi^{2}_\text{IP}$ and the transverse momenta of the $\Kstarz$ candidate children, the $\chi^{2}_\text{IP}$ and the transverse momenta of the $\D$ candidate children (only for two-body $\D$ final states), and the transverse momentum imbalance of the $\Bd$ candidate, defined as
    \begin{equation}
        I_{p_\text{T}}\equiv\frac{p_\text{T}\left(\Bd\right)-\sum_{X}p_\text{T}\left(X\right)}{p_\text{T}\left(\Bd\right)+\sum_{X}p_\text{T}\left(X\right)},
    \end{equation}
where the summation is over all charged tracks $X$ inconsistent with originating from the primary vertex within a cone around the $\Bd$ candidate, excluding those used in the $\Bd$ reconstruction~\cite{LHCb-PAPER-2020-036}. Variables related to the $\D$ children are omitted for the four-body $\D$ final states to avoid any dependence on the modelling of the $\D$ decay dynamics. Requirements on the output of each BDT classifier are chosen based on an optimisation of the estimated sensitivity to \CP-violating observables. Each BDT classifier retains roughly 85--90\% of the signal candidates and removes over $90\%$ of the combinatorial background.

Stringent particle identification (PID) requirements  are placed on the kaon child of the $\Kstarz$ candidate to reduce misidentification of the $\Kstarz$ flavour to negligible levels, as the charge of this particle is used to determine the flavour of the parent \B meson. More relaxed particle identification requirements are also placed on the pion from the $\Kstarz$ candidate and all kaons and pions from the two-body \D-meson decay candidates. For $\D\to \kaon\pion\pion\pion$ and $\D\to \pion \kaon\pion\pion$ decay candidates, the kaon and both pions with charge opposite to the $\D$ candidate kaon child are subject to PID requirements. For $\D\to\pion\pion\pion\pion$ candidates, only the two pions with the same charge as the $\Kstarz$ candidate kaon child are subject to PID requirements. All particles with PID requirements are also required to have track momentum between 3--200\gevc to ensure suitable discrimination between kaons and pions in the RICH detectors.

Requirements are also placed on the displacement of the $\D$ candidate decay vertex from the $\Bd$ candidate decay vertex to suppress charmless $\Bd$ decays to the same final states, which proceed without an intermediate $\D$ meson. Different requirements are placed on the ADS and GLW final states due to different relative contributions of charmless backgrounds. For the GLW (ADS) final states, the $\D$ candidate decay vertex displacement is required to be at least three (two) times its uncertainty. A discussion of the remaining contributions from these charmless backgrounds is included in Sect.~\ref{sec:backgrounds}.

Finally, vetoes are applied to each $\D$ final state to remove backgrounds or maintain consistency with the measurements of $\D$ hadronic parameters used in Sect.~\ref{sec:Interpretation}. In the ADS final states, $\D\to \kaon\pion$ decays misidentified as $\D\to\pion \kaon$ due to two incorrect PID assignments are removed with the requirement that the reconstructed $\D$ mass where the mass hypotheses of the kaon and pion children have been swapped differs from the known $\Dz$ meson mass by more than $15 \mevcc$. Backgrounds from $\Bpm \to \D\Kpm$ decays paired with a random pion are removed with a requirement that the $\D\Kpm$ reconstructed mass differ by more than $25 \mevcc$ from the known $\Bpm$ meson mass. Decays of the $\Bd$ meson to the same final state proceeding through different charmed intermediate states, \eg $\Bd\to \Dm[\Kp\pim\pim]\pip$ misidentified as $\Bd\to \D[\pip\pim]\Kp\pim$, are removed with requirements that the reconstructed mass of a $D$ child combined with the $\Kstarz$ candidate to form a Cabibbo-favoured decay of another charmed meson does not fall within $15 \mevcc$ of the known meson mass. Candidate $\D\to \KS\pip\pim$ decays are removed from the $\D\to \pip\pim\pip\pim$ sample with the requirement that the reconstructed $\pip\pim$ masses are not within 480--505\mevcc.  

The \CP-violating observables of interest are all proportional to the ratios of decay rates of different $\D$ final states, and thus much of the reconstruction and selection efficiency cancels in these ratios. However, due to the different kinematic distributions of each $\D$ final state, and due to different selection requirements placed on the different $\D$ decay modes, these efficiencies must be accounted for. Additionally, corrections for effects introducing asymmetries aside from those due to the interference discussed in Sect.~\ref{sec:Introduction} must be applied. 
The selection requirements are applied to simulated samples to estimate the selection efficiencies of each final state integrated across the two \B flavours, with the exception of PID efficiencies. Charge-dependent PID efficiencies are estimated with calibration data samples of kaons and pions collected from $\Dstar$ decays weighted to match the kinematics of signal decays \cite{LHCB-PUB-2016-021}. The resulting efficiency of reconstructing and selecting signal candidates is approximately $2\times 10^{-3}$.

Three other possible sources of flavour asymmetry must be accounted for, aside from that resulting from the interference discussed in Sect.~\ref{sec:Introduction}. First, an asymmetry can be introduced in the production of $\BdorBs$ mesons and their antiparticles from the initial proton-proton collisions, referred to as $A_\text{prod}$. Second, tracking and reconstruction can depend on the charge of final-state particles due to the difference of interactions between particles and anti-particles and the detector material, which is accounted for through the difference in detection asymmetry between kaons and pions $A_{\kaon\pion}=A_{\kaon}-A_{\pion}$. Lastly, similar to above, asymmetries can be introduced in PID which is accounted for implicitly in the PID efficiency corrections discussed previously. All discussed asymmetries are defined to match the flavour conventions introduced in Sect.~\ref{sec:Introduction}.

 The quantity $A_\text{prod}$ is determined from the measurement of $\Bd$ and $\Bs$ asymmetries in Ref.~\cite{LHCb-PAPER-2016-062}. The measured production asymmetries are $A_\text{prod}=\left(-8\pm 5\right)\times 10^{-3}$ and $A_{s,\text{prod}}=\left(6\pm 10\right)\times 10^{-3}$. Since these results are only determined using data from proton-proton centre-of-mass energies of $7\text{ and } 8\tev$, and not at $13\tev$, the same central values are assumed for the full dataset, but with inflated uncertainty. 

 The value of $A_{\kaon\pion}$ is estimated using calibration data samples of $\Dp\to \Km\pip\pip$ and $\Dp\to \KS \pip$ decays. The resulting predicted asymmetry weighted to match the signal kinematic distributions is $A_{\kaon\pion}=\left(-9.8\pm 5.5\right)\times 10^{-3}$. The total detection asymmetry for a $\Bd\to \D\Kstarz$ candidate depends on the difference between the number of kaons and the number of pions of the same sign in the final state, \eg the $\Bd\to \D[\kaon\pion]\Kstarz$ final state is corrected by $2A_{\kaon\pion}$, the $\Bd\to \D[\pion\pion]\Kstarz$ final state is corrected by $A_{\kaon\pion}$, while the $\Bd\to \D[\pion \kaon]\Kstarz$ final state requires no such correction.

\section{\texorpdfstring{Measurement of \CP-violating observables}{Measurement of CP-violating observables}}
\label{sec:Fit}
The \CP-violating observables discussed in Sect.~\ref{sec:Introduction} are determined from the selected data through a simultaneous unbinned extended maximum-likelihood fit of the $\Bd$  reconstructed mass for each flavour of each $\D$ final state, where constraints are placed on the $\D$ candidate mass and the direction of the \B candidate momentum using the \textsc{DecayTreeFitter} package \cite{Hulsbergen:2005pu}. The reconstructed mass is calculated with these constraints unless otherwise specified. Only candidates with a reconstructed mass within the interval 5000--5800\mevcc are considered. 

\subsection{Modelling the reconstructed mass distribution}
\label{sec:backgrounds}

The reconstructed mass distributions of both $\Bd\to \D\Kstarz$ and $\Bs\to \D\Kstarzb$ decays are modelled by modified Cruijff functions\cite{LHCb-PAPER-2020-019} parameterised as
     \begin{equation}
	f(M)=\begin{cases}
	 e^{\frac{-\left(M-\mu\right)^2\left(1+
				\beta\left(M-\mu\right)^2\right)}{2\sigma^2+\alpha_L\left(M-\mu\right)^2}},&M<\mu \\
		 e^{\frac{-\left(M-\mu\right)^2\left(1+
				\beta\left(M-\mu\right)^2\right)}{2\left(\rho \sigma\right)^2+\alpha_R\left(M-\mu\right)^2}},& M>\mu 
	\end{cases}.
\end{equation}
For each decay mode, all the parameters of the $\BdorBs\to \D\Kstarz$ modified Cruijff functions are fixed from fits to the corresponding simulation samples of $\Bd\to \D[\kaon\pion]\Kstarz$ and $\Bs\to \D[\pion \kaon]\Kstarz$ decays, with the exception of the mean, $\mu$, and width, $\sigma$,  which are left as free parameters. The constraints on the $\D$ meson mass minimise the dependence on the $\D$ decay final state, but the four-body $\D$ final states have a slightly larger width. The ratio of the two-body width and four-body width is determined from simulation and fixed in the fit to data.

The selection requirements accept $\BdorBs\to \Dstar\Kstarz$ decays with a similar efficiency to signal decays, as their topology is identical to signal decays with one additional missing neutral particle from the $\Dstar\to \gamma/\piz \D$ decay. The reconstructed masses of these decays are mainly located below the $\Bd$ mass peak, but the right-hand tail of the $\Bs \to \Dstar\Kstarzb$ distribution extends to the $\Bd$ meson mass. As this is the decay of a pseudoscalar particle into two vector particles, the final state is characterised by three independent helicity amplitudes (0, $\pm1$). However, two of these amplitudes ($\pm1)$ have indistinguishable reconstructed mass distributions. This leaves four different configurations of $\BdorBs\to \Dstar\Kstarz$ decays: two possible \Dstar decay chains each with two reconstructed mass distributions. Each configuration is modelled with either a single broad peaking structure or a double-peaked structure, as described in Ref.~\cite{LHCb-PAPER-2020-036}. The relative size of the $\Dstar[\gamma \D]$ and $\Dstar[\piz \D]$ decays for each helicity state are fixed based on efficiencies predicted from simulation and the measured branching fractions. The interference effects of interest can affect the helicity amplitudes differently. Therefore, four free parameters are included in the fit to account for this: one for the ratio of helicity states in ADS $\Bd\to \Dstar\Kstarz$ decays, one ratio for each flavour of the GLW $\Bd\to \Dstar\Kstarz$ decays, and one ratio shared across all $\Bs\to \Dstar\Kstarzb$ final states.

   Due to particle misidentification, $\Bd\to \D\pip\pim$ and $\Bd\to \Dstar\pip\pim$ decays can also be mistaken for signal decays when one pion is misidentified as a kaon. The relatively stringent PID requirement on the $\Kp$ child of the $\Kstarz$ candidate suppresses these contributions significantly, but not to negligible levels. The reconstructed mass distribution of these decays is taken from simulated samples with PID weights determined from calibration data samples. The simulation samples only consider the contribution of $\Bd\to \D\rhoz[\pip\pim]$ and $\Bd\to \Dstar\rhoz[\pip\pim]$, as the selection requirements on the $\Kstarz$ candidate significantly suppress other contributions. The reconstructed mass distribution of $\Bd\to \D \rhoz[\pip\pim]$ decays is modelled with a double-sided Crystal Ball function\cite{Skwarnicki:1986xj}. The probability distribution function (PDF) for $\Bd\to \Dstar
    \rhoz[\pip\pim]$ decays is parameterised in a similar fashion to the $\mbox{\Bd\to \Dstar \Kstarz}$ decays, and the PDF shares the ratio of helicity states with ADS $\Bd\to \Dstar\Kstarz$ decays. The peak of the $\Bd\to D
    \rhoz[\pip\pim]$ distribution is located between the $\Bd$ and $\Bs$ peaks, while the $\Bd\to \Dstar
    \rhoz[\pip\pim]$ distribution has minimal contribution near or above the $\Bd$ mass peak.

    Partially reconstructed $\Bu\to \D\Kp\pip\pim$ decays, where the $\pip$ is not included in the reconstruction, also pass final selection requirements. The reconstructed mass distribution of these decays is modelled with a smoothed kernel density estimation\cite{RooKeysPDF} of simulated $\Bu\to\D \kaon_1(1270)^+$ decays, and contributes negligibly near the $\Bd$ mass peak, being distributed primarily around a reconstructed mass of $5050\mevcc$.

Requirements placed on the $\D$ candidate flight significance do not fully suppress the contribution of charmless $\Bd$ and $\Bs$ decays to the same final states as the observed signal channels. The size of these contributions and possible flavour asymmetries in each final state are estimated using data in the sidebands of the $\D$-candidate reconstructed mass distribution. However, as the default selection and reconstruction algorithm applies a mass constraint on the $D$ candidate, the $\Bd$-candidate reconstructed-mass distribution in those sidebands is biased. To address this, the number of charmless decays in our selection sample is estimated with samples from the $D$-candidate sideband regions for each final state  by fitting to the $\Bd$ candidate reconstructed-mass distribution without any constraint applied on the reconstructed $D$-candidate mass. The number of events found in the sidebands are used to estimate the number of charmless decays that pass the selection requirements. The magnitude of the flavour asymmetry in these decays is estimated with a similar procedure, however with a significantly less stringent requirement on the $\D$ flight significance to increase statistical precision. Significant charmless contributions from $\Bs$ decays are only observed in the $\D\to\pion \kaon$ and $\D\to \pion \kaon\pion\pion$ final states. The largest contribution from charmless decays is found in the $\Bd\to
\D[\pi\pi]\Kstarz$ and $\Bdb\to\D[\pi\pi]\Kstarzb$ samples with a yield that is roughly 10\% of the determined signal yield.

The combinatorial background is modelled by an exponential PDF. The parameters of the exponential PDF associated with each $\D$ decay mode vary freely in the fit, but are shared across $\B$ flavour for each $\D$ decay mode.

\subsection{Parameterisation of the reconstructed mass distribution}
\label{sec:fitsetup}
The fit to determine the number of signal decays in each of the fourteen samples (two flavours for each of the seven $\D$ final states) is parameterised in terms of the \CP-violating observables introduced in Sect.~\ref{sec:Introduction}, the flavour-integrated number of observed $\Bd\to \D[\kaon\pion]\Kstarz$ decays, and the flavour-integrated number of observed \linebreak$\Bd\to \D[\kaon\pion\pion\pion]\Kstarz$ decays. These observables are related to the number of observed decays in each signal mode and the efficiency and asymmetry corrections discussed in Sect.~\ref{sec:backgrounds}. The $\mathcal R^{hh(\pion\pion)}_{CP}$ parameters are dependent on measured $\Dz$ branching fractions, reported in Table~\ref{table:InputBFS}.

\begin{table}[tb]
	\centering{}
 	\caption{Summary of branching fractions used as inputs in the measurement.}\label{table:InputBFS}
	\begin{tabular}{llc}
		\hline
		Branching Fraction&Value\\\hline
		$\mathcal B\left(\Dz\to \Km\pip\right)$&$\left(3.999\pm0.045\right)\%$&\cite{HeavyFlavorAveragingGroup:2022wzx}\\
		$\mathcal B\left(\Dz\to \pip\pim\right)$&$\left(1.490\pm0.027\right)\times 10^{-3}$&\cite{HeavyFlavorAveragingGroup:2022wzx}\\
		$\mathcal B\left(\Dz\to \Kp\Km\right)$&$\left(4.113\pm0.051\right)\times 10^{-3}$&\cite{HeavyFlavorAveragingGroup:2022wzx}\\
		$\mathcal B\left(\Dz\to \Km\pip\pip\pim\right)$&$\left(8.22\phantom{0}\pm0.14\right)\%$&\cite{PDG2022}\\
		$\mathcal B\left(\Dz\to \pip\pim\pip\pim\right)$&$\left(8.06\phantom{0}\pm0.20\right)\times 10^{-3}$&\cite{PDG2022}\\ \hline
	\end{tabular}
\end{table}

The $\Bs\to D\Kstarzb$, $\Bd\to \Dstar\Kstarz$, $\Bs\to \Dstar\Kstarzb$, and $\Bu\to \D\Kp\pip\pim$ background components are parameterised in a similar fashion. The \CP-violating observables of $\Bs\to \D\Kstarzb$ decays are left as free parameters in the fit, as are the \CP-violating observables of $\Bd\to \Dstar\Kstarz$ decays. The observables of $\Bs\to \Dstar\Kstarzb$ decays are fixed to have no effects from interference, and the \CP-violating observables of $\Bu\to \D\Kp\pip\pim$ decays are fixed to the measurements for the two-body $\D$ final states from Ref.~\cite{LHCb-PAPER-2015-020} and predicted values based on the results of Ref.~\cite{LHCb-CONF-2022-003} for the four-body $\D$ final states. The $\Bd\to \D\pip\pim$ and $\Bd\to \Dstar\pip\pim$ yields are constrained to be equal for $\Bd\to \D[ \kaon\pion]\Kstarz$ and $\Bd\to D[\pion \kaon]\Kstarz$ decays,  due to an equal probability of misidentifying the $\pip$ meson and the $\pim$ meson. For all other decay modes, the yields are constrained relative to the yields in $\Bd\to \D[\kaon\pion]\Kstarz$, and scaled by the ratio of $\D$ branching fractions and selection efficiencies. The number of charmless decays is subject to Gaussian constraints according to the determined values. The combinatorial backgrounds are parameterised in terms of the observed number of candidates, which is required to be equal for each flavour of a given $\D$ decay mode. The possibility of biases introduced in the fitting procedure is investigated by fitting toy data samples generated from the result of the total PDF fit to data. No evidence is seen for biases in the central values or uncertainties of any of the \CP-violating observables. 

\subsection{Systematic uncertainties}
\label{sec:systematics}
The following sources of systematic uncertainties are considered: uncertainty in the determination of the asymmetry and efficiency corrections, uncertainty from the measured branching fractions used as inputs in the fit, and uncertainty in the modelling of the PDFs or in the determination of the fixed yields in the fit. Any systematic uncertainties that are estimated at less than $5\%$ of the measured statistical uncertainty from the fit are considered negligible, and are excluded in the final determination of systematic uncertainties. A summary of the assigned systematic uncertainties is provided in Tables~\ref{table:B0Systematics} and \ref{table:BsSystematics}. The total systematic uncertainty is calculated as the sum in quadrature of the individual sources. The dominant systematic uncertainties are due to the measurement of input branching fractions, asymmetry corrections, and estimation of particle identification efficiencies, but are generally small compared to statistical uncertainties. The assigned systematic uncertainties are slightly larger than those in the previous analysis from Ref.~\cite{LHCb-PAPER-2019-021} as more detailed study is possible with the larger dataset.

The uncertainties of $A_\text{prod}$ and $A_{\kaon\pion}$ are propagated to the parameters of interest based on the quoted uncertainties in Sect.~\ref{sec:Selection}. The measured uncertainty of $A_\text{prod}$ \cite{LHCb-PAPER-2016-062} is doubled for the data samples collected at $13 \tev$ to account for possible energy dependence in the asymmetry. Additionally, the possibility of asymmetries introduced by the hardware trigger is estimated to be at the level of $10^{-3}$ based on the results of Ref.~\cite{LHCb-PAPER-2020-036}.  The uncertainty due to the selection efficiencies largely cancels, with the primary uncertainty arising from different PID requirements on different final states. Uncertainties in estimated PID efficiencies arise from the limited statistics of simulated samples for kinematic weighting and the finite size of bins used in the weighting of the PID efficiency. The propagated uncertainties of the input branching fractions listed in Table~\ref{table:InputBFS} are also determined.

The uncertainty due to the modelling of the $\Bd$ candidate reconstructed mass for each component in the fit is also investigated. The uncertainty due to the modelling of $\Bs\to \Dstar \Kstarz$ decays is assessed by varying the determined efficiencies and input branching fractions of the $\Dstar[\gamma \D]$ and $\Dstar[\piz \D]$ components within their respective uncertainties. Two alternative PDFs are examined to assess uncertainty in the modelling of $\Bu\to \D\Kp\pip\pim$ decays: one based on simulated samples of $\Bu\to \D\Kstar(1400)^0[\Kp\pip\pim]$ decays, and another based on $\Bu\to \D\Kp\pip\pim$ decays collected from data. Performing the fit with either distribution results in negligible variations in the \CP-violating observables. An alternative $\Bd\to \D\pip\pim$ reconstructed mass distribution composed of an equal admixture of $\Bd\to \D\rhoz[\pip\pim]$ and non-resonant $\Bd\to \D\left(\pion\pion\right)_{S-\text{wave}}$ decays based on results in Ref.\cite{LHCb-PAPER-2014-070} is also considered. These uncertainties are negligible for all \CP-violating observables except $R^{\pm}_{\pi K(\pi\pi)}$.

The statistical uncertainty of the flavour-integrated number of charmless decays is  accounted for through Gaussian constraints in the fit, but the asymmetry of the charmless yields is fixed based on the central values determined in data.  The resulting uncertainty on the parameters of interest is determined through variations of the charmless asymmetries within their measured uncertainties.

The assumption of equal combinatorial background in $\Bz$ and $\Bzb$ candidate samples is studied using pseudoexperiments, where an ensemble of datasets is generated with asymmetric background yields. The magnitude of the asymmetry generated is determined from an alternative fit to the data. The pseudoexperiments are fit with the default model and biases in the fitted observables are taken as the systematic uncertainty.

\begin{landscape}
	\begin{table}[p]
		\caption{Summary of uncertainties on the physics observables measured in $\Bz\to \D \Kstarz$ decays. Uncertainties smaller than $5\%$ of the statistical uncertainty are indicated with dashed lines.}\label{table:B0Systematics}
		\centering{}
		\begin{tabular}{lcccccccccccc}
			\hline
			& $\mathcal A_{\kaon\pion}$ & $\mathcal R^{+}_{\pion \kaon}$&$\mathcal R^{-}_{\pion \kaon}$ & $\mathcal R^{\pion\pion}_{CP}$ & $\mathcal{A}^{\pion\pion}_{CP}$& $\mathcal R^{\kaon\kaon}_{CP}$ & $\mathcal{A}^{\kaon\kaon}_{CP}$&$\mathcal A_{\kaon\pion\pion\pion}$ & $R^{+}_{\pion \kaon\pion\pion}$&$\mathcal R^{-}_{\pion \kaon\pion\pion}$ & $\mathcal R^{4\pion}_{CP}$ & $\mathcal{A}^{4\pion}_{CP}$\\ \hline
			$A_\text{prod}$&$0.009$& $0.001$ & $0.001$ & $0.002$& $0.009$ & --& $0.009$ & $0.009$ & $0.001$&$0.001$& --& $0.009$\\
			$A_{\kaon\pion}$&$0.009$& $0.001$ & $0.001$ & $0.001$& $0.006$ & $0.001$& $0.006$ & $0.009$ & $0.001$&$0.001$& $0.001$& $0.006$\\
			PID &$0.008$& $0.004$ & $0.004$ & $0.012$& $0.010$ & $0.010$& $0.010$ & $0.010$ & $0.004$&$0.004$& $0.017$& $0.011$\\
    Fit PDFs& --&	0.003&	0.003&	--&	--&	--&	--&	--&	0.003	&0.003 &	--&	--\\
        Charmless Asymmetries& --&	--&	--&	0.011&	0.005&	--&	--&	--&	0.001&	0.001 &	--&	--\\
        Combinatorial Asymmetries& --&	0.002& 0.002&	--&	0.002&	--&	0.001&	--&	0.004 &	0.004 &	--&	0.001\\
			Input Branching Fractions& --& --& --& 0.020	& --&	0.013 &	--& --&	--&	--&	0.028&	--\\
			\hline
			Total systematic&$0.015$& $0.005$& $0.005$ & $0.026$& $0.016$ & $0.017$& $0.015$ & $0.016$ & $0.006$&$0.006$& $0.033$& $0.016$\\
			\hline
			Statistical&$0.017$& $0.013$& $0.013$ & $0.110$& $0.103$ & $0.057$& $0.064$ & $0.018$ & $0.014$& $0.014$& $0.084$& $0.088$
		\end{tabular}
	\end{table}

 	\begin{table}[p]
		\caption{Summary of uncertainties on the physics observables measured in $\Bs\to \D \Kstarz$ decays. Uncertainties smaller than $5\%$ of the statistical uncertainty are indicated with dashed lines.}\label{table:BsSystematics}
		\centering{}
		\begin{tabular}{lcccccccccccc}
			\hline
			& $\mathcal A_{s,\pi K}$ & $\mathcal R^{+}_{s,K \pi}$&$\mathcal R^{-}_{s, K\pi}$ & $\mathcal R^{s,\pi\pi}_{CP}$ & $\mathcal{A}^{s,\pi\pi}_{CP}$& $\mathcal R^{s,KK}_{CP}$ & $\mathcal{A}^{s,KK}_{CP}$&$\mathcal A_{s,\pi K\pi\pi}$ & $R^{+}_{s,K\pi\pi\pi}$&$\mathcal R^{-}_{s,K \pi \pi\pi}$ & $\mathcal R^{s,4\pi}_{CP}$ & $\mathcal{A}^{s,4\pi}_{CP}$\\ \hline
			$A_{s,prod}$&$0.018$& $0.003$ & $0.003$ & --& $0.018$ & --& $0.018$ & $0.018$ & $0.003$&$0.003$& --& $0.018$\\
			$A_{K\pi}$&---& $0.002$ & $0.002$ & $0.001$& $0.006$ & $0.001$& $0.006$ &  --& $0.002$&$0.002$& $0.001$& $0.006$\\
			PID &$0.008$& $0.004$ & $0.004$ & $0.012$& $0.010$ & $0.010$& $0.010$ & $0.010$ & $0.004$&$0.004$& $0.017$& $0.011$\\
    Fit PDFs& --&	0.002&	0.002&	--&	--&	--&	--&	--&	0.002 &	0.002 &	--&	--\\
        Charmless Asymmetries& --&	--&	--&	--&	--&	--&	--&	--&	--&	-- &	--&	--\\
           Combinatorial Asymmetries& --&	0.002& 0.002&	--&	0.002&	--&	0.001&	--&	0.004 &	0.004 &	--&	0.001\\
			Input Branching Fractions& --& --& --& 0.020	& --&	0.013 &	--& --&	--&	--&	0.028&	--\\
			\hline
			Total Systematic&$0.020$& $0.006$& $0.006$ & $0.023$& $0.021$ & $0.016$& $0.021$ & $0.021$ & $0.007$&$0.007$& $0.033$& $0.022$\\
			\hline
			Statistical&$0.011$& $0.002$& $0.002$ & $0.056$& $0.057$ & $0.032$& $0.034$ & $0.012$ & $0.004$& $0.004$& $0.044$& $0.048$
		\end{tabular}
	\end{table}
\end{landscape}

\subsection{Results of the reconstructed mass fit}
\label{sec:results}

\begin{figure}[htbp]
	\centering{}
	\begin{tabular}{cc}
		\includegraphics[width=8.0cm]{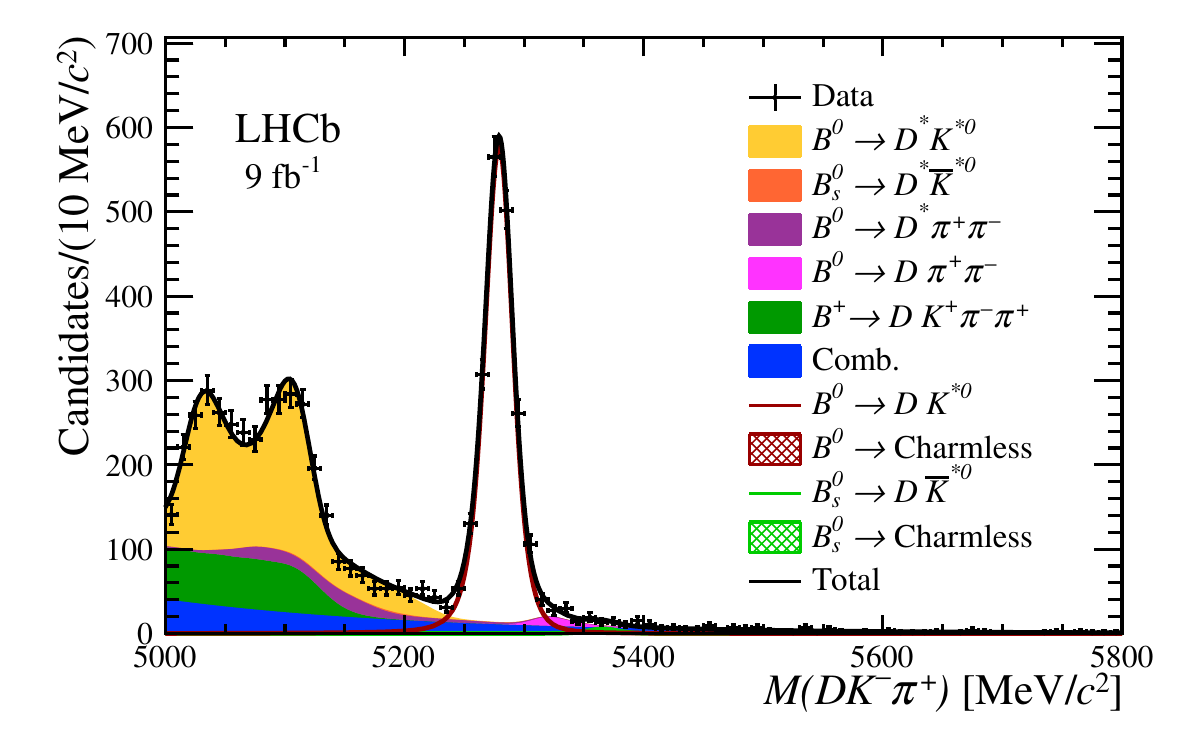}&\includegraphics[width=8.0cm]{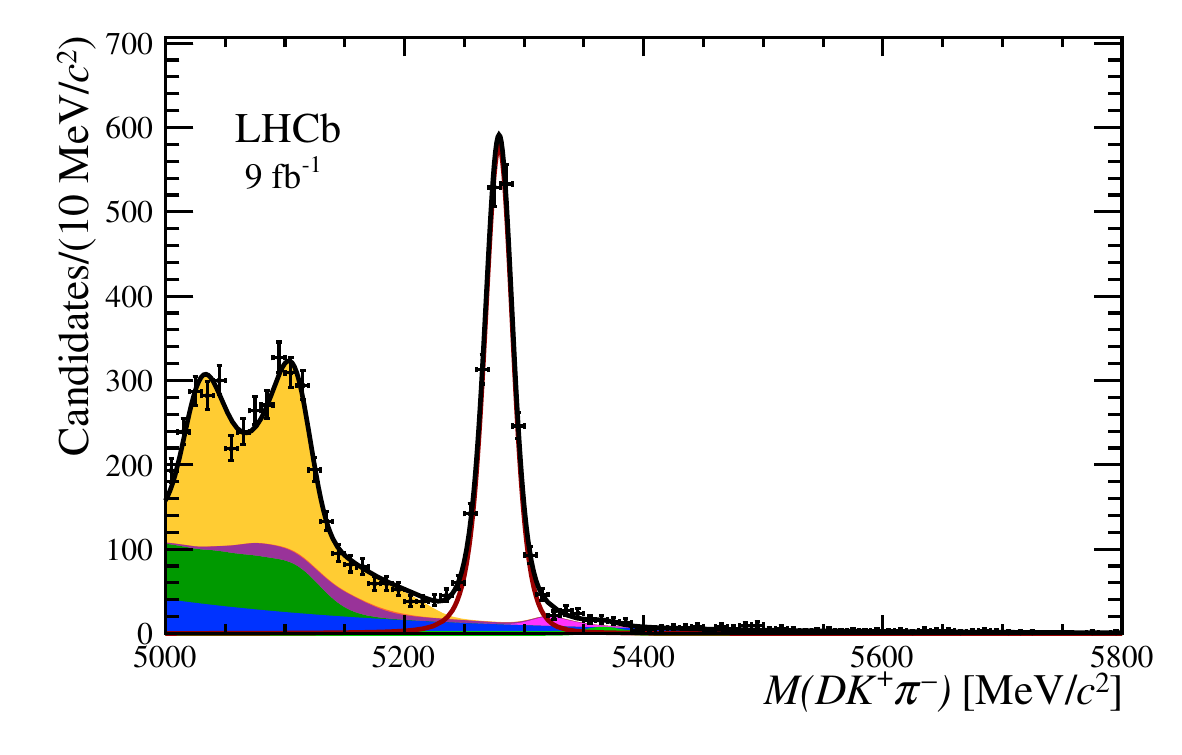}		\\
  \includegraphics[width=8.0cm]{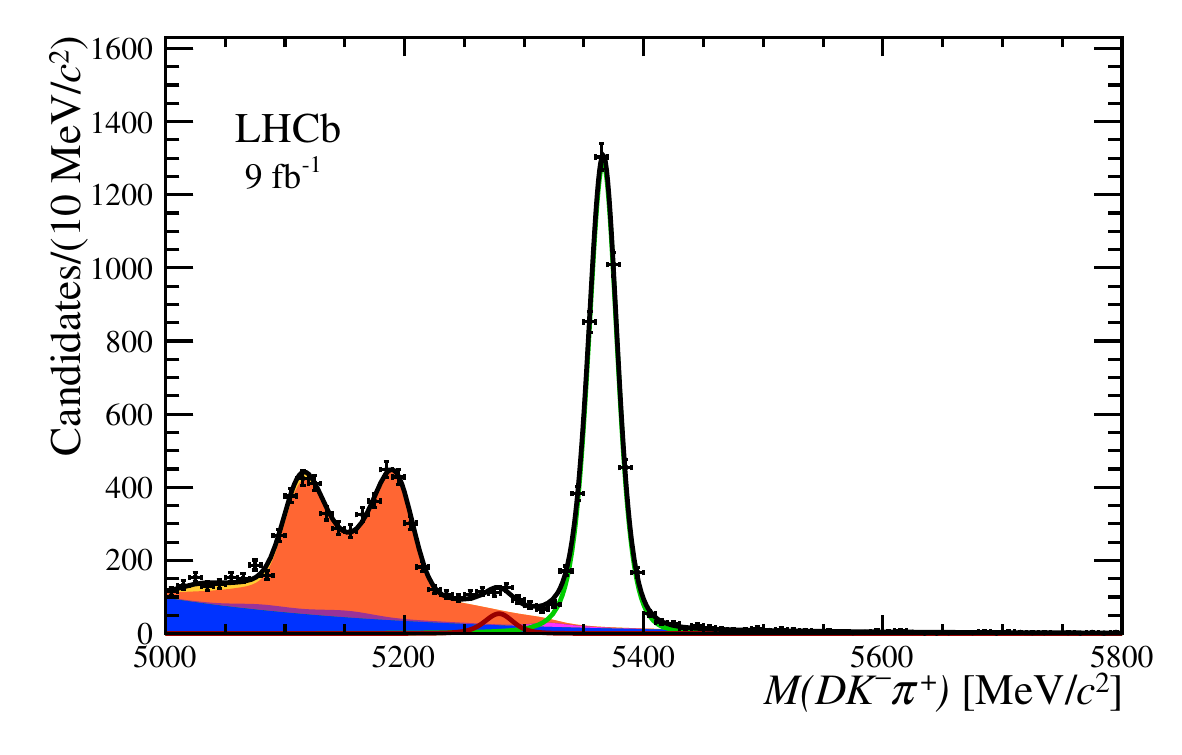}&\includegraphics[width=8.0cm]{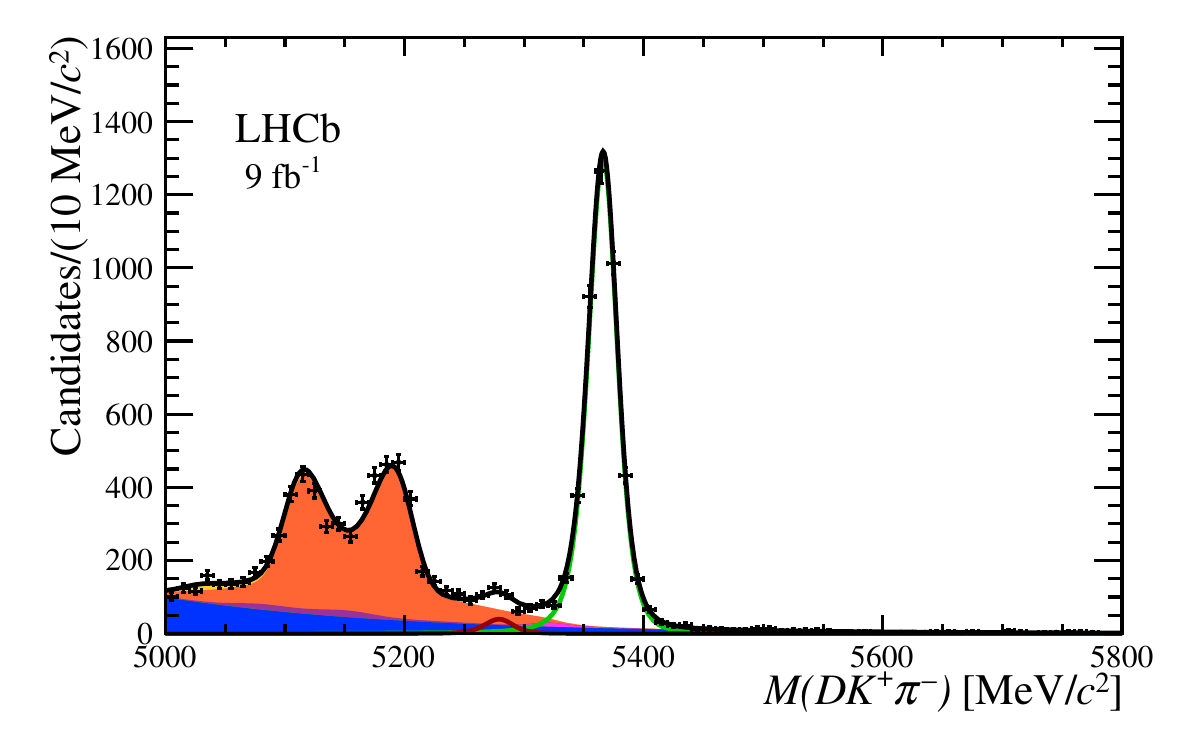}
	\end{tabular}
	
	\caption{Reconstructed mass distributions for selected candidates in the (left) $\Bdb$ and (right) $\Bd$ samples for the (top) $\D\to\kaon\pion$ and (bottom) $\D\to\pion\kaon$ decay channels. The fit projection is overlaid.}\label{fig:KPiFit}
\end{figure}

\begin{figure}[htbp]
	\centering{}
	\begin{tabular}{cc}
		\includegraphics[width=8.0cm]{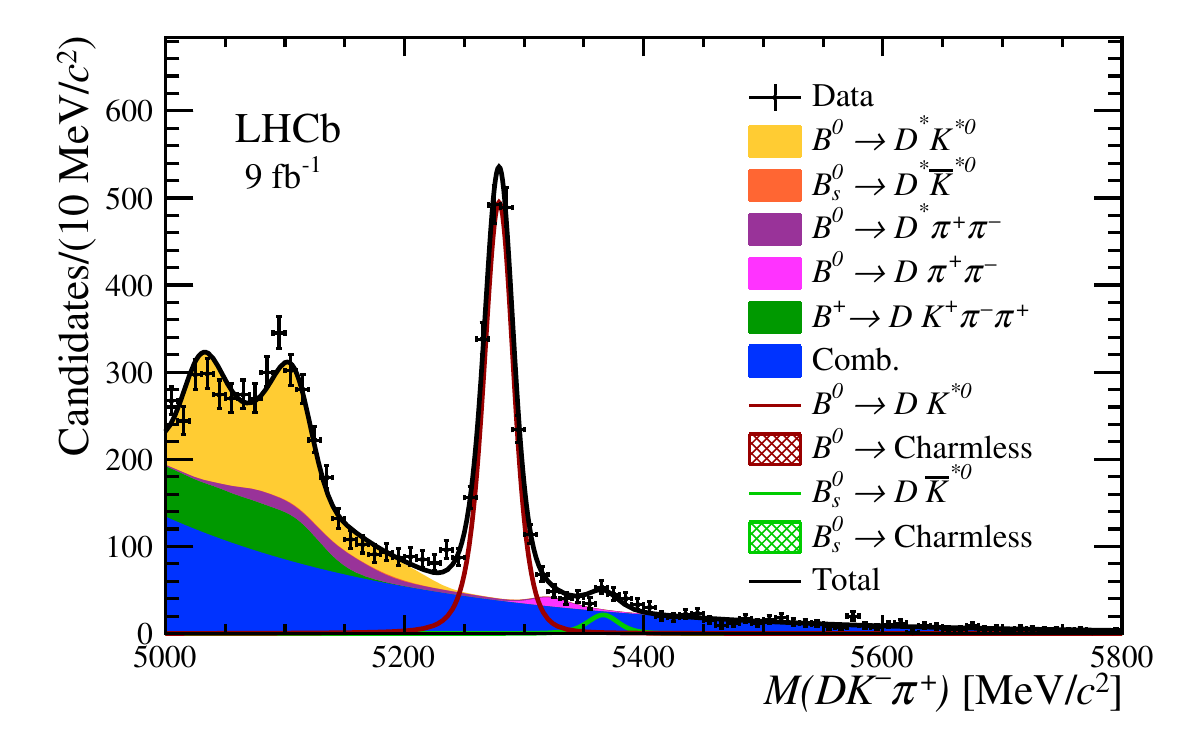}&\includegraphics[width=8.0cm]{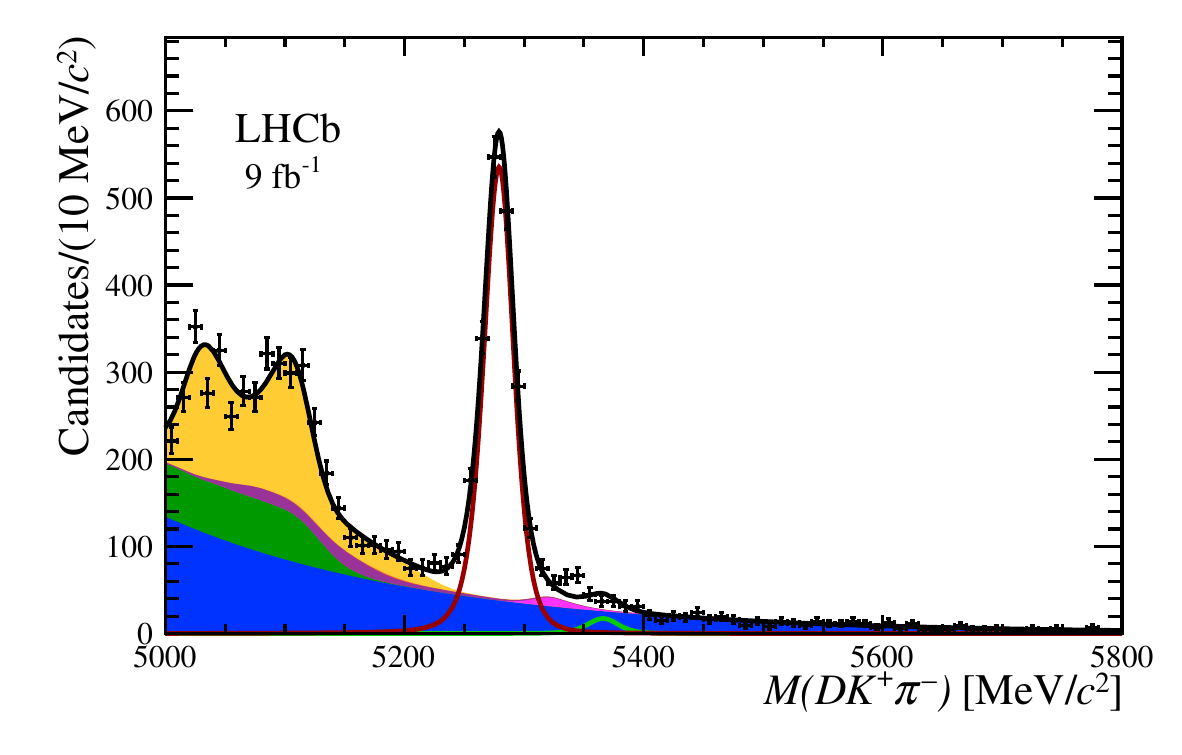}\\
  	\includegraphics[width=8.0cm]{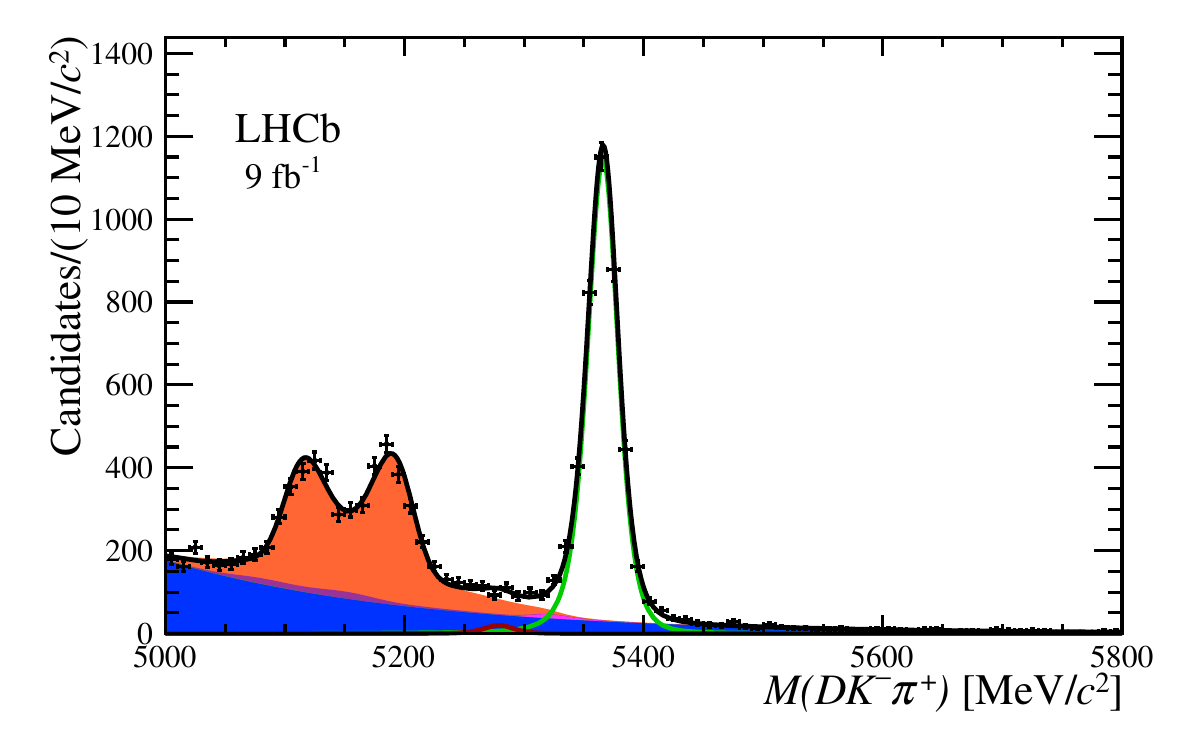}&\includegraphics[width=8.0cm]{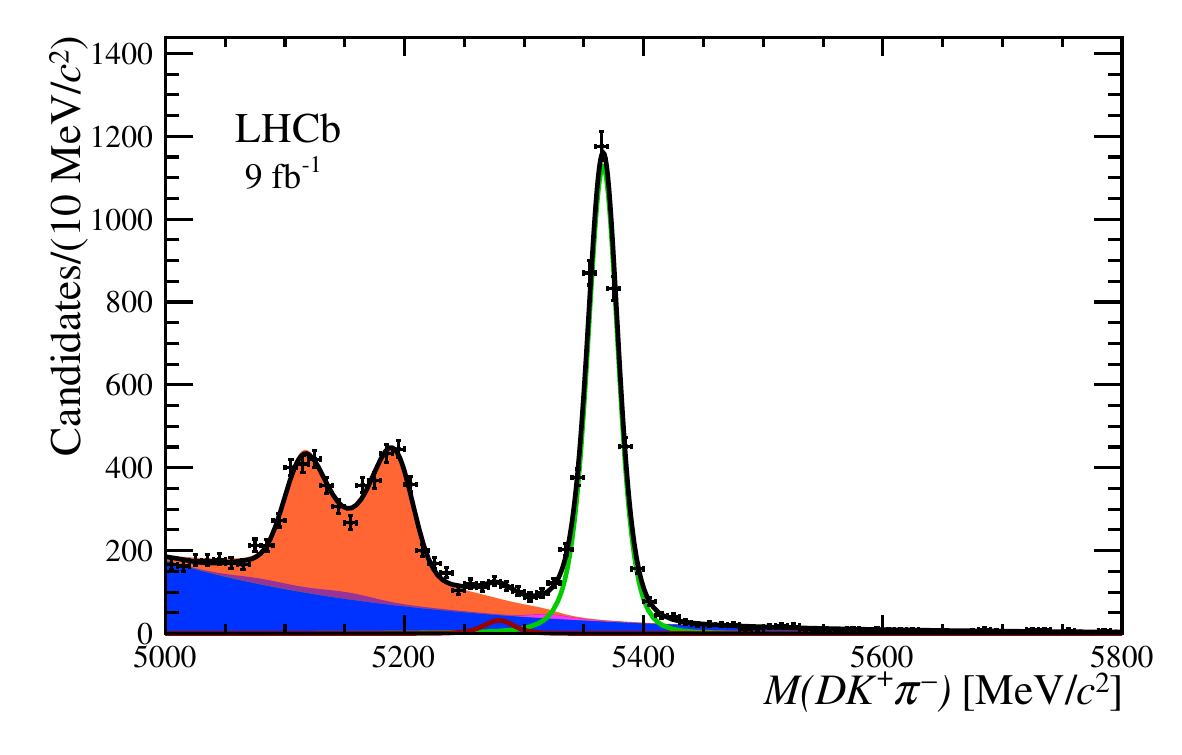}	
	\end{tabular}
	
	\caption{Reconstructed mass distributions for selected candidates in the (left) $\Bdb$ and (right) $\Bd$ samples for the (top) $\D\to\kaon\pion\pion\pion$ and (bottom) $\D\to\pion\kaon\pion\pion$ decay modes. The fit projections are overlaid.}\label{fig:K3PiFit}
\end{figure}

\begin{figure}[htbp]
	\centering{}
	\begin{tabular}{cc}
		\includegraphics[width=8.0cm]{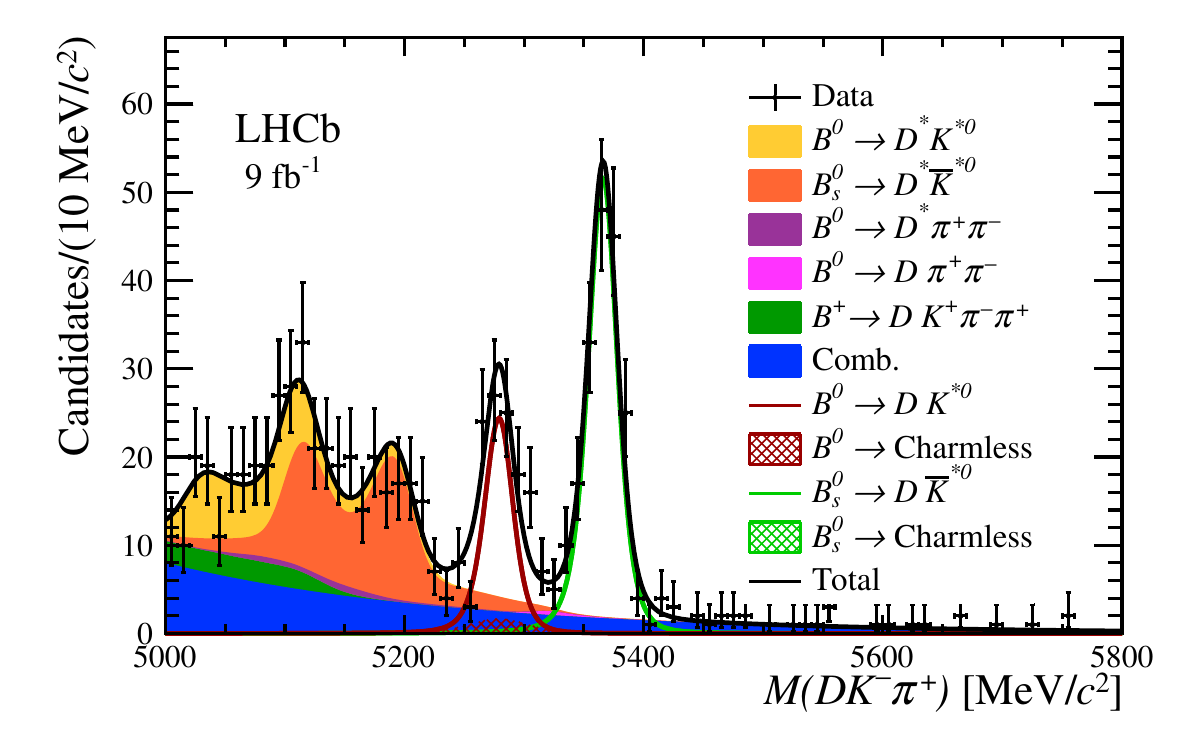}&\includegraphics[width=8.0cm]{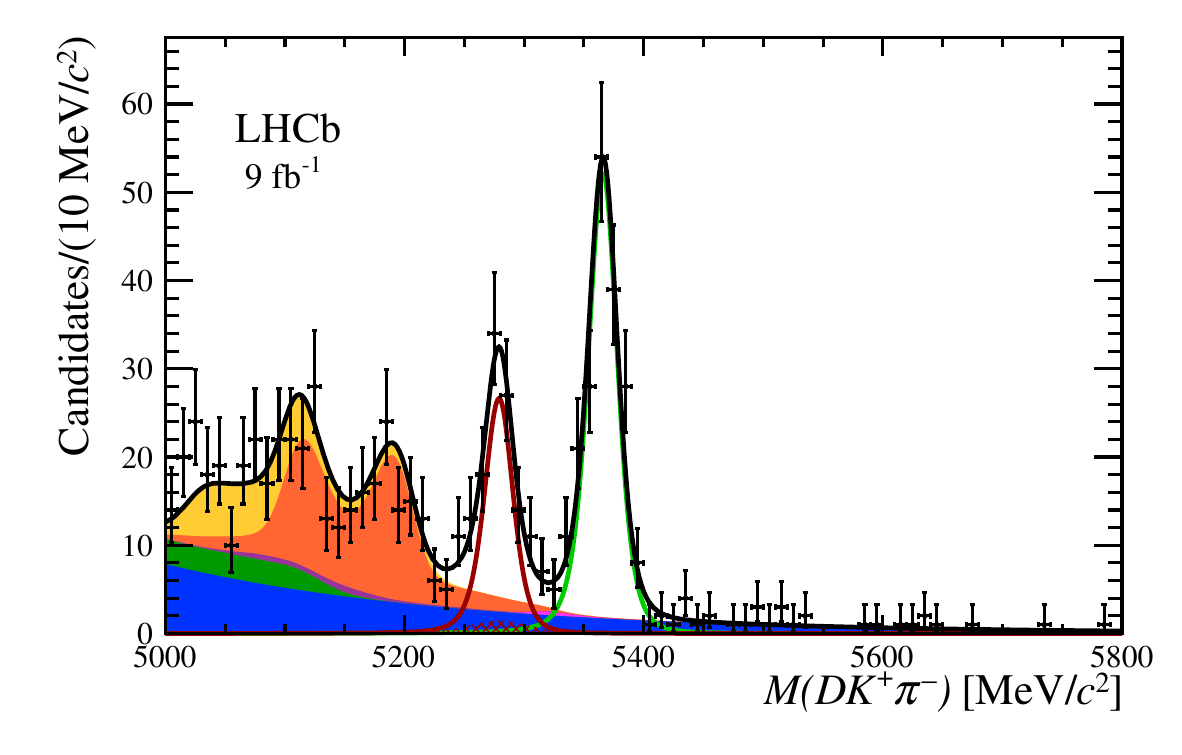}\\
  	\includegraphics[width=8.0cm]{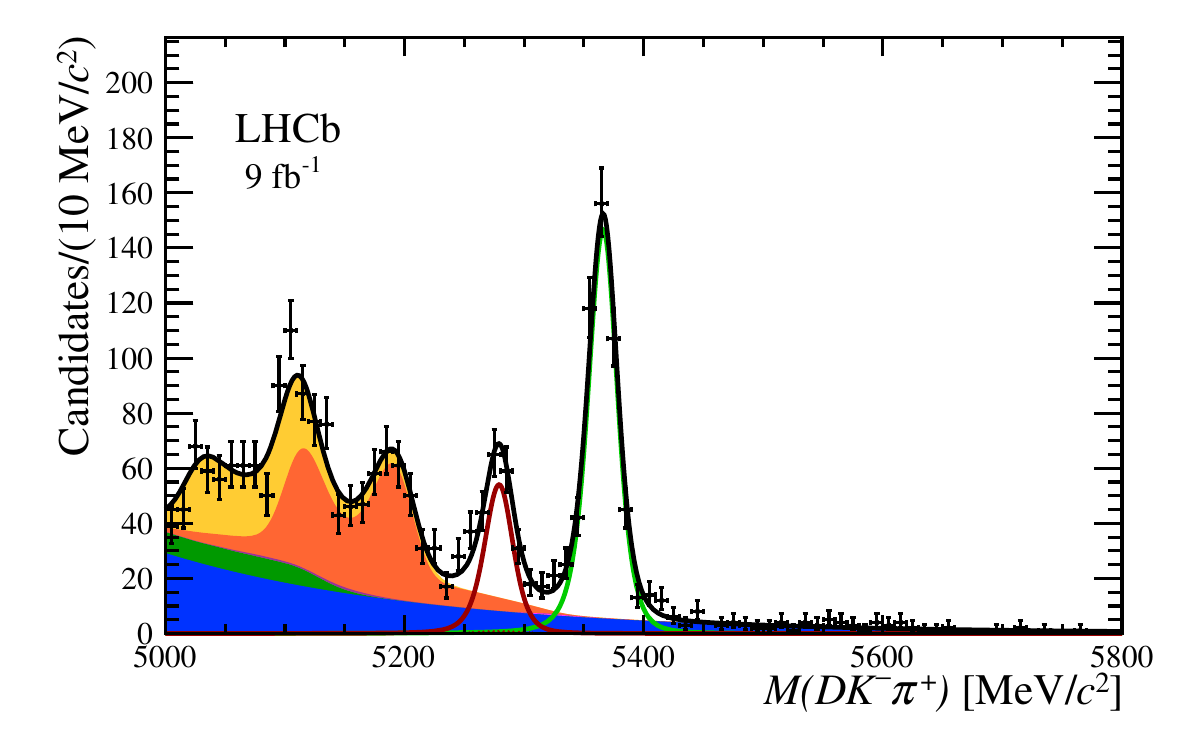}&\includegraphics[width=8.0cm]{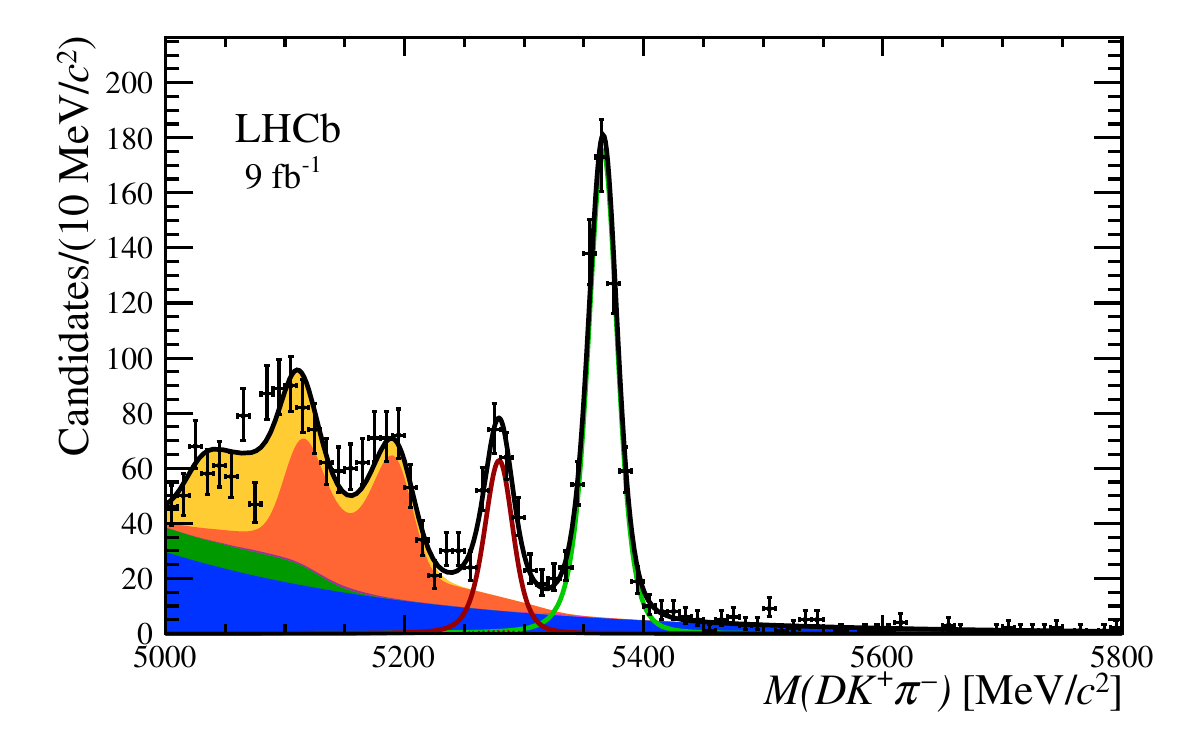}\\
   	\includegraphics[width=8.0cm]{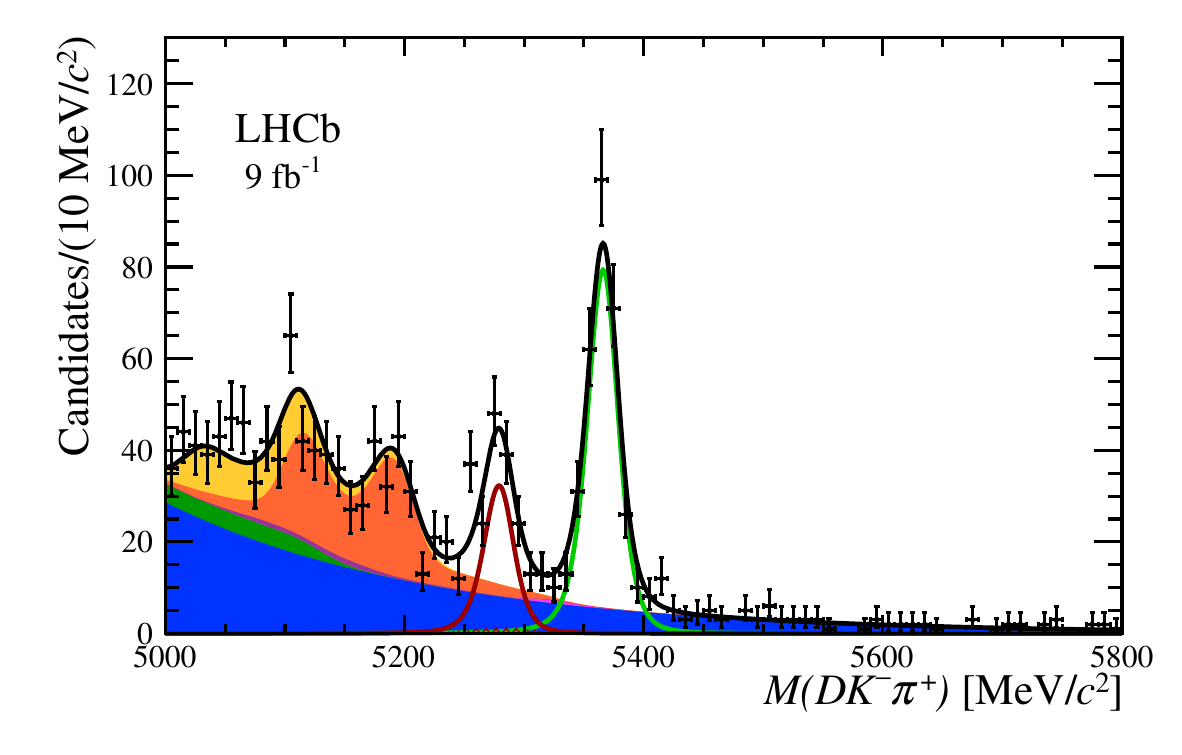}&\includegraphics[width=8.0cm]{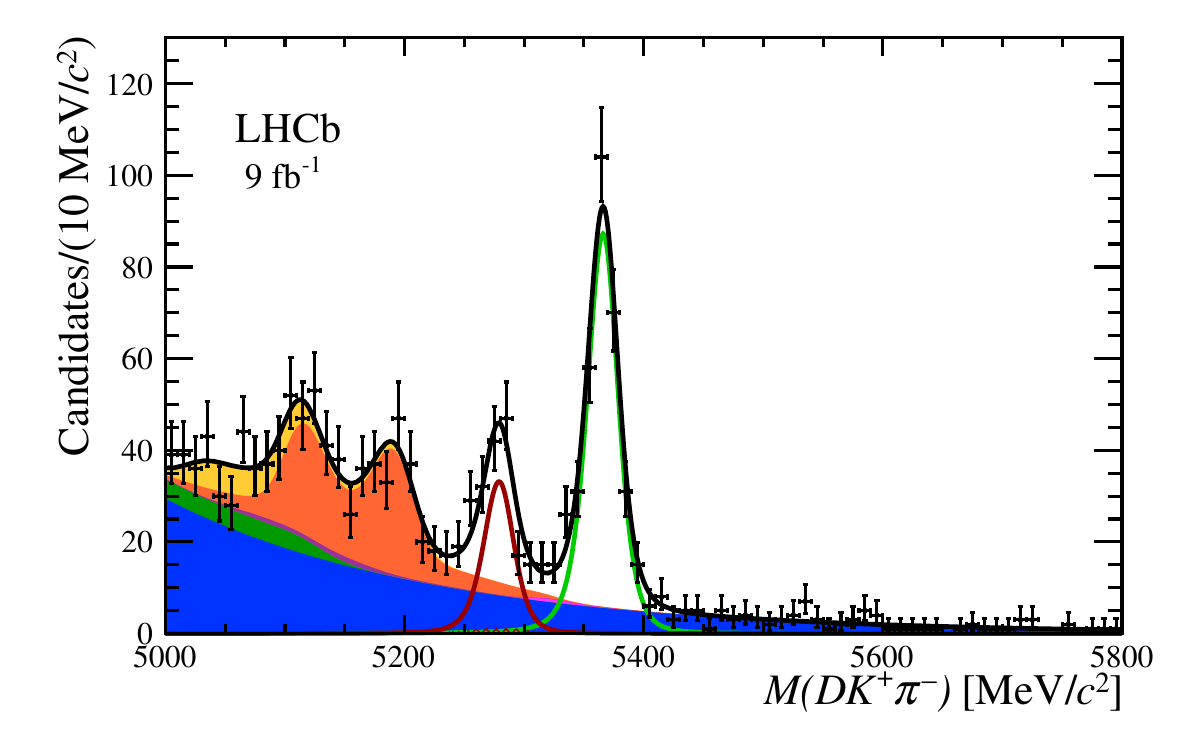}
		
	\end{tabular}
	
	\caption{Reconstructed mass distributions for selected candidates in the (left) $\Bdb$ and (right) $\Bd$ samples for the (top) $\D\to\pion\pion$, (centre) $\D\to\kaon\kaon$, and (bottom) $\D\to\pion\pion\pion\pion$ decay modes. The fit projections are overlaid.}\label{fig:GLWFit}
\end{figure}

The reconstructed \B candidate mass distributions for different \D decay modes are shown with the results of the fit overlaid in Figs.~\ref{fig:KPiFit}--\ref{fig:GLWFit}. The yields of the favoured decays, $N_{K\pi},\;N_{K\pi\pi\pi},N_{s,\pi K},\text{ and } N_{s,\pi K\pi\pi}$ are found to be roughly 3800, 3600, 8800, and 8200, respectively. The \CP-violating observables determined in the fit for $\mbox{\Bd\to \D \Kstarz}$ decays are given in Table~\ref{table:B0FitVals} and the results of $\Bs \to \D \Kstarzb$ are given in Table~\ref{table:BsFitVals}. The results from the $\Bs$ decays conform to the expectation of unobservable effects of interference: all $\mathcal R^s_{CP}$ parameters are consistent with unity within one standard deviation, and no statistically significant asymmetries are observed. However, statistically significant evidence of interference is seen in the $\Bd\to \D \Kstarz$ results, most notably in $\mathcal R^{\kaon\kaon}_{CP}$ at the $3\sigma$ level and $\mathcal R^{4\pi}_{CP}$ at the $2\sigma$ level. An approximately $2\sigma$ tension is observed between the values of $\mathcal R^{\kaon\kaon}_{CP}$ and $\mathcal R^{\pion\pion}_{CP}$, which are expected to be consistent, but this difference is attributed to a statistical fluctuation. The asymmetries $\mathcal A^{\kaon\kaon}_{CP}$ and $\mathcal A^{\pion\pion}_{CP}$ are consistent with each other. In aggregate, the results are self-consistent, as evidenced by our interpretation of the parameters in Sect.~\ref{sec:Interpretation}. 

Statistical precision on the \CP-violating observables has improved by around 60$\%$ in comparison to the previous results in Ref.~\cite{LHCb-PAPER-2019-021}, an improvement which is consistent with the increase in signal yield. Central values are also consistent taking into account correlations and the more accurate determination of the contribution of charmless decays with the larger dataset.

\begin{table}[hbtp]
	\centering{}
 	\caption{Fitted \CP-violating observables related to $\Bd\to D \Kstarz$ decays. The first stated uncertainties are statistical and the second are systematic.}\label{table:B0FitVals}
	\npdecimalsign{.}
	\nprounddigits{4}
	\begin{tabular}{cc}
		\hline
		Parameter & Value\\
		\hline
		$\mathcal A_{\kaon\pion}$ & $   \phantom{-}0.031   \pm   0.017 \pm 0.015 $\\ 
		$\mathcal R_{\pion \kaon}^+$ & $   \phantom{-}0.069  \pm   0.013 \pm 0.005  $\\ 
		$\mathcal R_{\pion \kaon}^-$ & $   \phantom{-}0.093   \pm   0.013 \pm 0.005$\\ 
		$\mathcal A_{\kaon\pion\pion\pion}$ & $   -0.012  \pm   0.018 \pm 0.016  $\\ 
		$\mathcal R_{\pion \kaon\pion\pion}^+$ & $  \phantom{-}0.060\pm   0.014 \pm 0.006 $\\ 
		$\mathcal R_{\pion \kaon\pion\pion}^-$ & $   \phantom{-}0.038\pm   0.014 \pm 0.006$\\ 
		$\mathcal R^{\kaon\kaon}_{CP}$ & $   \phantom{-}0.811   \pm   0.057 \pm 0.017$\\ 
		$\mathcal A^{\kaon\kaon}_{CP}$ & $   -0.047  \pm   0.063 \pm 0.015$\\ 
		$\mathcal R^{\pion\pion}_{CP}$ & $   \phantom{-}1.104   \pm   0.111 \pm 0.026$\\ 
		$\mathcal A^{\pion\pion}_{CP}$ & $   -0.034  \pm   0.094 \pm 0.016$\\ 
		$\mathcal R^{4\pion}_{CP}$ & $   \phantom{-}0.882 \pm   0.086 \pm 0.033$\\ 
		$\mathcal A^{4\pion}_{CP}$ & $   \phantom{-}0.021  \pm   0.087 \pm 0.016$\\ 
		\hline
	\end{tabular}
\end{table}

\begin{table}[hbtp]
	\centering{}
 	\caption{Fitted \CP-violating observables related to $\Bs\to \D \Kstarzb$ decays. The first stated uncertainties are statistical and the second are systematic.}\label{table:BsFitVals}
	\npdecimalsign{.}
	\nprounddigits{4}
	\begin{tabular}{cc}
		\hline
		Parameter & Value\\
		\hline
		$\mathcal A_{s,\pion \kaon}$ & $        -0.009 \pm 0.011 \pm 0.020    $  \\ 
		$\mathcal R_{s,\kaon\pion}^+$ & $       \phantom{-}0.004 \pm 0.002   \pm 0.006   $  \\ 
		$\mathcal R_{s,\kaon\pion}^-$ & $       \phantom{-}0.004 \pm 0.002 \pm 0.006     $  \\ 
		$\mathcal A_{s,\pion \kaon\pion\pion}$ & $  -0.029 \pm 0.012 \pm 0.021    $  \\ 
		$\mathcal R_{s,\kaon\pion\pion\pion}^+$ & $ \phantom{-}0.019 \pm 0.004 \pm 0.007     $  \\ 
		$\mathcal R_{s,\kaon\pion\pion\pion}^-$ & $ \phantom{-}0.015 \pm 0.004  \pm 0.007     $  \\ 
		$\mathcal R^{s,\kaon\kaon}_{CP}$ & $      \phantom{-}1.000 \pm 0.034 \pm 0.016   $  \\ 
		$\mathcal A^{s,\kaon\kaon}_{CP}$ & $      \phantom{-}0.062 \pm 0.032  \pm 0.021    $  \\ 
		$\mathcal R^{s,\pion\pion}_{CP}$ & $ \phantom{-}  0.996 \pm 0.057   \pm 0.023   $  \\ 
		$\mathcal A^{s,\pion\pion}_{CP}$ & $  -0.001\pm 0.056   \pm 0.021   $  \\ 
		$\mathcal R^{s,4\pion}_{CP}$ & $   \phantom{-} 1.010 \pm 0.048    \pm 0.033  $  \\ 
		$\mathcal A^{s,4\pion}_{CP}$ & $    \phantom{-}0.017 \pm 0.044    \pm0.022  $  \\ 
		\hline
	\end{tabular}
\end{table}

\section{Interpretation and conclusions}
\label{sec:Interpretation}

The measured parameters from Table~\ref{table:B0FitVals} can be expressed in terms of the fundamental physics parameters of interest, $\gamma$, $r_{\Bd}^{\D\Kstar}$, and $\delta_{\Bd}^{\D\Kstar}$, and some additional inputs. These additional inputs include the coherence factor $\kappa_\Bd$, which quantifies the dilution of the interference effects of interest due to the selected $\Bd\to \D\Kp\pim$ decays that do not proceed through an intermediate $\Kstarz$ resonance. The value of this parameter measured by Ref.~\cite{LHCb-PAPER-2015-059} is used, which placed identical requirements on the $\Kstarz$ candidate to those presented in this article. There is additional dependence on the hadronic parameters associated with the $\D$ meson decays to a final state $X$: $r_\D^{X}$, $\delta_\D^{X}$, and $\kappa_\D^{X}$ for ADS decays, and the \CP-even fraction $F_+^{X}$ for GLW decays. The values of the ADS parameters are taken from Ref.~\cite{LHCb-CONF-2022-003} for the $\D\to \kaon\pion$ decay (except for $\kappa_\D^{K\pi}$, which is exactly one)  and from Ref.~\cite{BESIIIK3Pi} for the $\D\to \kaon\pion\pion\pion$ decay. For the two-body GLW modes, $F_+^{hh}$ is exactly one, and the value of $F_+^{4\pion}=0.746\pm0.013$ is taken from an average of the results from Ref.~\cite{BESIIIFPlus4Pi} and Ref.~\cite{CLEOFPlus4Pi}. The GLW observables $\mathcal A_{CP}^{hh(\pi\pi)}$ and $\mathcal R_{CP}^{hh(\pi\pi)}$ relate to these parameters through
\begin{equation}
	\begin{array}{r@{}l}
		\mathcal A^{hh(\pion\pion)}_{CP}&=\frac{2\kappa_\Bd r_{\Bd}^{\D\Kstar}\left(2F_+^{hh(\pion\pion)}-1\right)\sin\left(\delta_{\Bd}^{\D\Kstar}\right)\sin\left(\gamma\right)}{1+\left(r_{\Bd}^{\D\Kstar}\right)^2+2\kappa_\Bd r_{\Bd}^{\D\Kstar} \left(2F_+^{hh(\pion\pion)}-1\right)\cos\left(\delta_{\Bd}^{\D\Kstar}\right)\cos\left(\gamma\right)}\text{,}\\
		\\
		\mathcal R^{hh(\pion\pion)}_{CP} &=\frac{1+\left(r_{\Bd}^{\D\Kstar}\right)^2+2\kappa_\Bd r_{\Bd}^{\D\Kstar}\left(2F_+^{hh(\pion\pion)}-1\right)\cos\left(\delta_{\Bd}^{\D\Kstar}\right)\cos\left(\gamma\right)}{1+\left(r_{\Bd}^{\D\Kstar}\right)^2\left(r_\D^{\kaon \pion \pion\pion}\right)^2+2\kappa_\Bd r_{\Bd}^{\D\Kstar} r_\D^{\kaon \pion \pion\pion}\kappa_{D}^{\kaon \pion\pion\pion }\cos\left(\delta_{\Bd}^{\D\Kstar}-\delta_\D^{\kaon \pion\pion\pion}\right)\cos\left(\gamma\right)}.
	\end{array}
  \label{eq:GLWEqs}
\end{equation}

The $\mathcal R^+_{\pion\kaon(\pion\pion)}$ and $\mathcal R^-_{\pion\kaon(\pion\pion)}$ observables can be expressed, neglecting effects of charm mixing, as
\begin{equation} \small
	\mathcal R^\pm_{\pion\kaon(\pion\pion)} =\frac{\left(r_{\Bd}^{\D\Kstar}\right)^2+\left(r_\D^{\kaon \pion (\pion\pion)}\right)^2+2\kappa_\Bd r_{\Bd}^{\D\Kstar} r_\D^{\kaon \pion (\pion\pion)}\kappa_{D}^{\kaon \pion(\pion\pion) }\cos\left(\delta_{\Bd}^{\D\Kstar}+\delta_\D^{\kaon \pion (\pion\pion)}\pm\gamma\right)}{1+\left(r_{\Bd}^{\D\Kstar}\right)^2\left(r_\D^{\kaon \pion (\pion\pion)}\right)^2+2\kappa_\Bd r_{\Bd}^{\D\Kstar} r_\D^{\kaon \pion (\pion\pion)}\kappa_{D}^{\kaon\pion(\pion\pion) }\cos\left(\delta_{\Bd}^{\D\Kstar}-\delta_\D^{\kaon \pion (\pion\pion)}\pm\gamma\right)}.
	\label{eq:Rpm}
\end{equation}
 Additional corrections are made due to the effects of charm mixing, as described in Ref.~\cite{CharmMixing}, which are of the order of $1\%$. Finally, the parameters $\mathcal A_{\kaon\pion}$ and $\mathcal A_{\kaon\pion\pion\pion}$ can be expressed as
\begin{equation}
\small
 \mathcal A_{\kaon\pion(\pion\pion)}=\frac{2\kappa_\Bd\kappa_\D^{\kaon\pion(\pion\pion)}r_\D^{\kaon\pion(\pion\pion)}\sin\left(\delta_{\Bd}^{\D\Kstar}-\delta_\D^{\kaon\pion(\pion\pion)}\right)\sin(\gamma)}{1+\left(r_{\Bd}^{\D\Kstar}\right)^2\left(r_\D^{\kaon\pion(\pion\pion)}\right)^2+2\kappa_\Bd\kappa_\D^{\kaon\pion(\pion\pion)}r_\D^{\kaon\pion(\pion\pion)}\cos\left(\delta_{\Bd}^{\D\Kstar}-\delta_\D^{\kaon\pion(\pion\pion)}\right)\cos(\gamma)}.
 	\label{eq:AKPi}
\end{equation}

\begin{figure}[htbp]
	\centering{}
 \begin{tabular}{cc}
		\includegraphics[width=8.0cm]{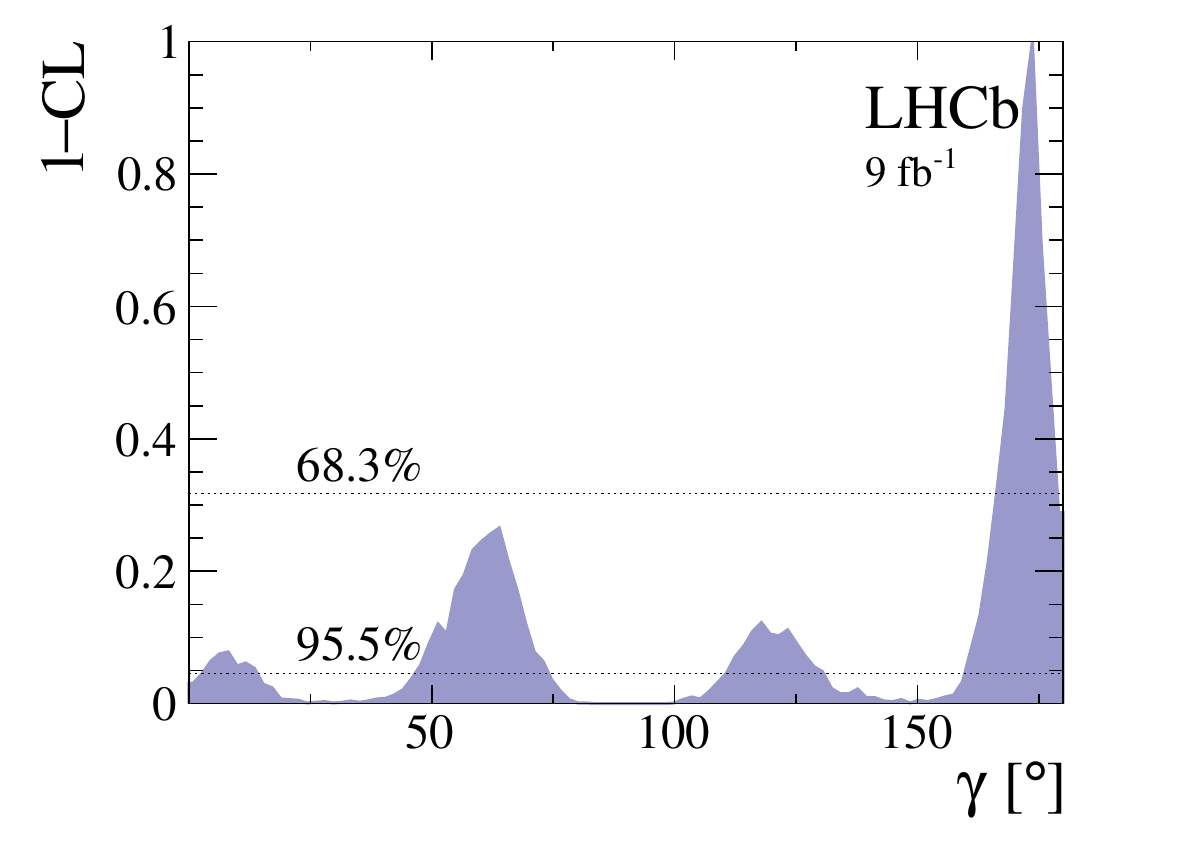}&\includegraphics[width=8.0cm]{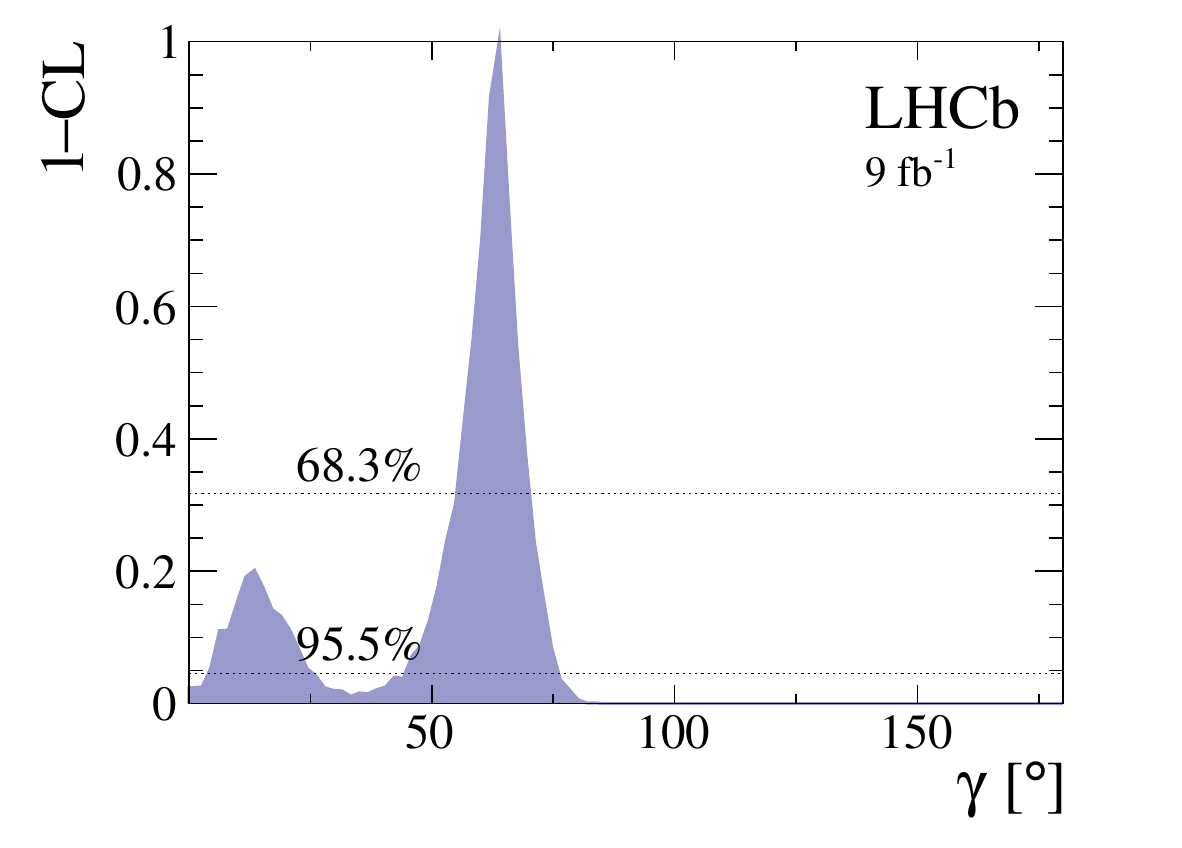}
  \end{tabular}
	
	\caption{Confidence levels from a frequentist analysis of the $\Bd$-related \CP-violating observables from  (left) just this analysis and (right) this analysis combined with the results from Ref.~\cite{LHCb-PAPER-2023-009} on the CKM angle $\gamma$.}\label{fig:gammaOnly}
\end{figure}

\begin{figure}[h!]
	\centering{}
	\begin{tabular}{cc}
		\includegraphics[width=8.0cm]{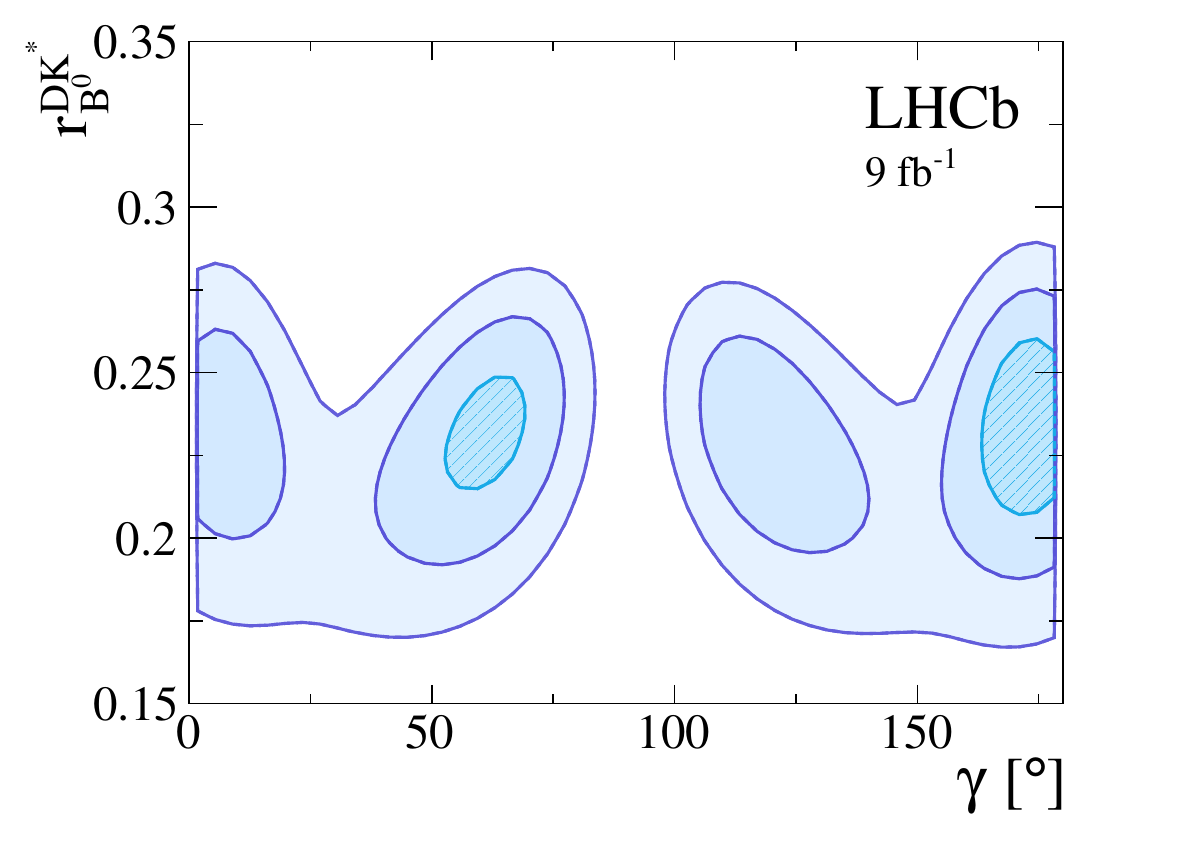}&\includegraphics[width=8.0cm]{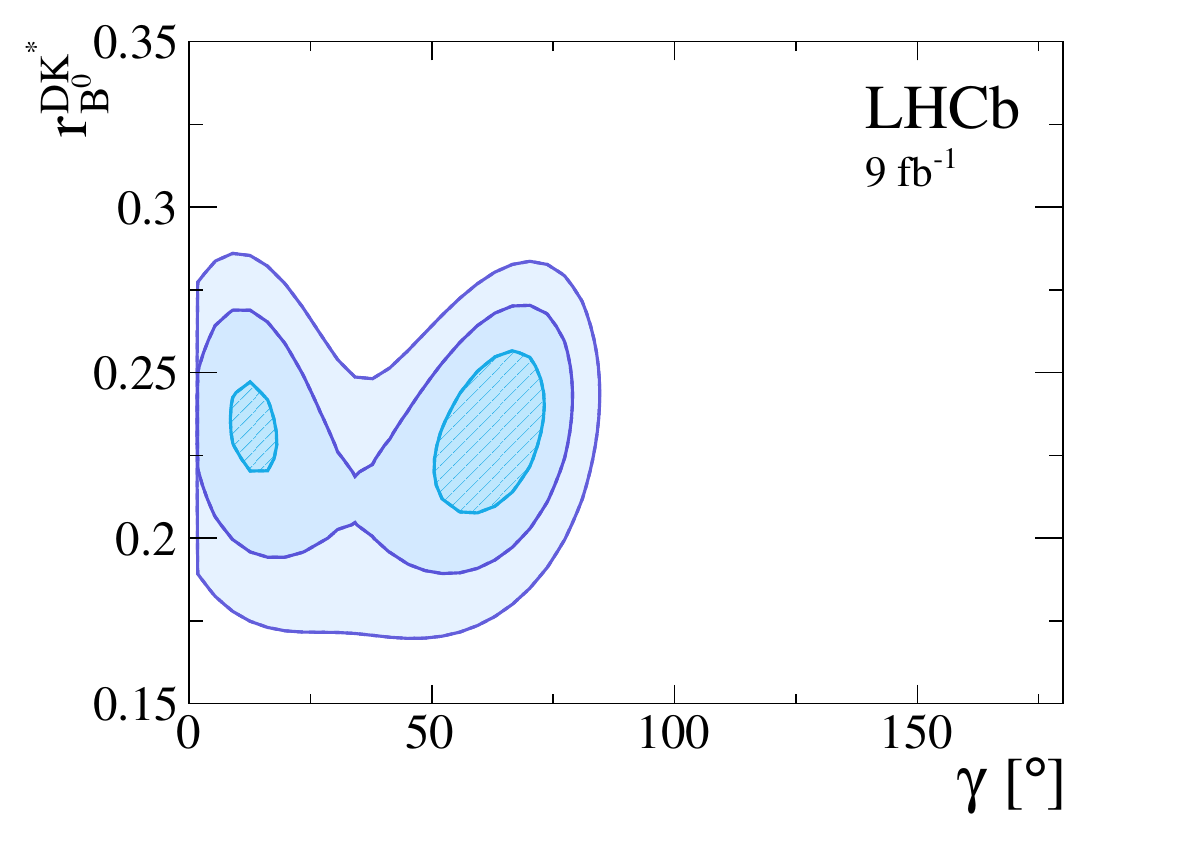}
	\end{tabular}
	
	\caption{Confidence level contours from a profile-likelihood analysis of the $\Bd$-related \CP-violating observables from  (left) just this analysis and (right) this analysis combined with the results from Ref.~\cite{LHCb-PAPER-2023-009} projected to the $\gamma$-$r_{\Bd}^{\D\Kstarz}$ plane. Contours contain 68.3\%, 95.4\%, and 99.7\% of the distribution.}\label{fig:gammaRb}
\end{figure}

\begin{figure}[h!]
	\centering{}
	\begin{tabular}{cc}
		\includegraphics[width=8.0cm]{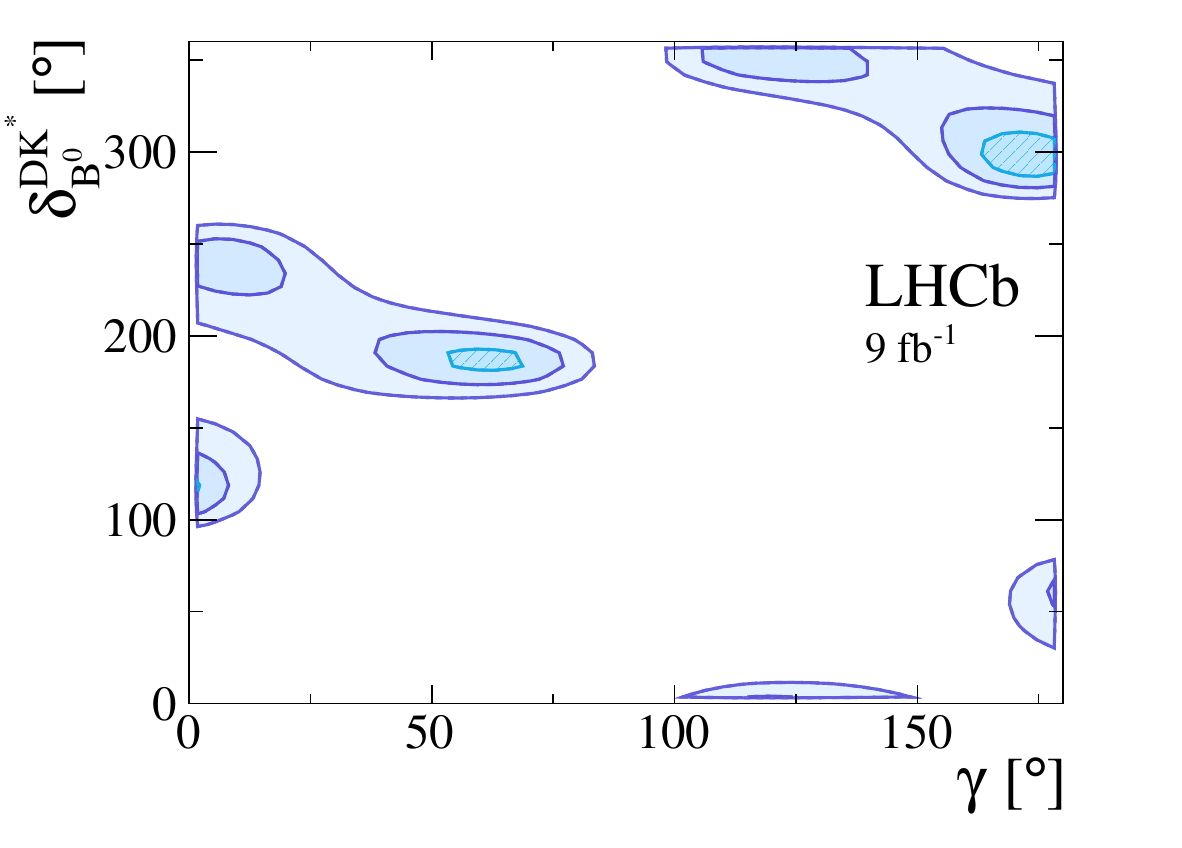}& 	\includegraphics[width=8.0cm]{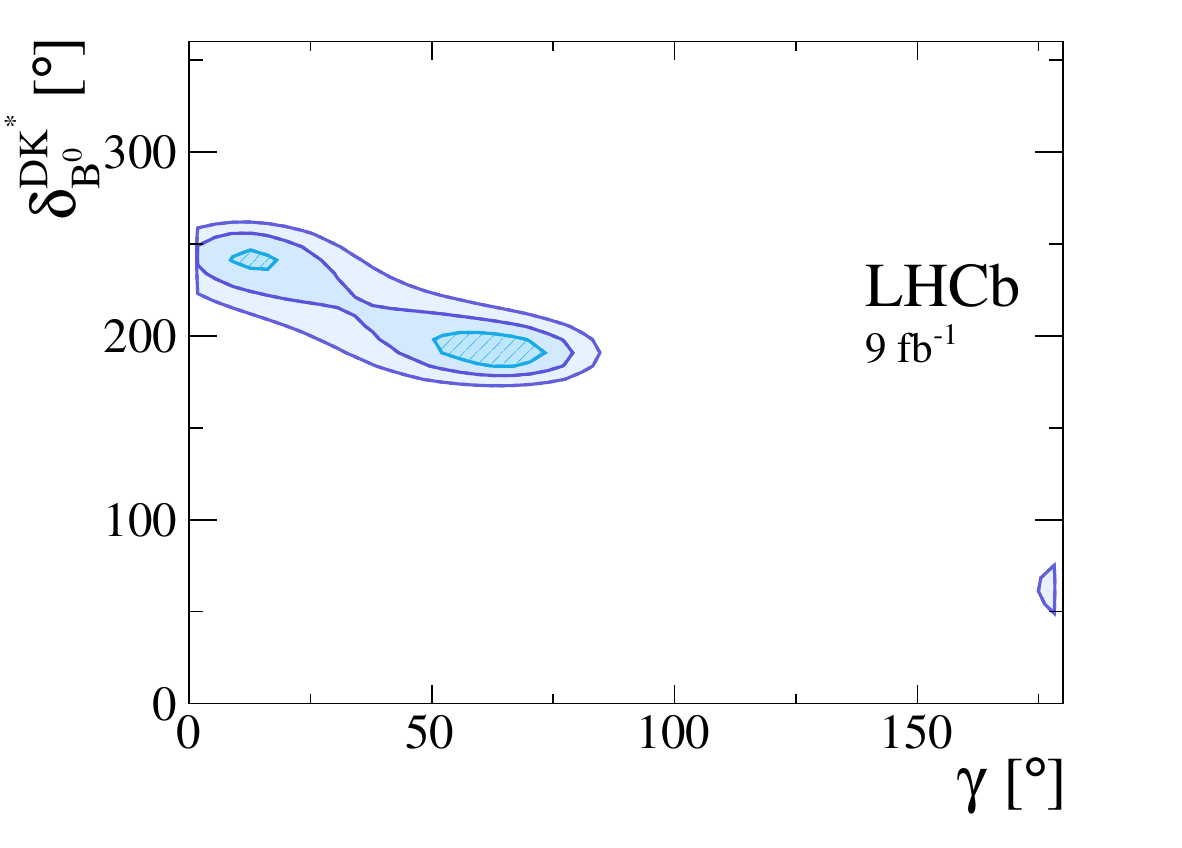}	
	\end{tabular}
	
	\caption{Confidence level contours from a profile-likelihood analysis of the \CP-violating observables from  (left) just this analysis and (right) this analysis combined with the results from Ref.~\cite{LHCb-PAPER-2023-009} projected to the $\gamma$-$\delta_\Bd^{\D\Kstarz}$ plane. Contours contain 68.3\%, 95.4\%, and 99.7\% of the distribution.}\label{fig:gammaDb}
\end{figure}

With Eqs.~\ref{eq:GLWEqs}--\ref{eq:AKPi}, the measured \CP-violating observables from Table~\ref{table:B0FitVals}, and the aforementioned inputs,  limits are set on the ($\gamma$, $r_{\Bd}^{\D\Kstar}$, $\delta_{\Bd}^{\D\Kstar}$) parameter space using the GammaCombo package \cite{LHCb-PAPER-2016-032}, which can implement both profile-likelihood and frequentist methods. The correlations of statistical and systematic uncertainties for the measured \CP-violating observables are accounted for with correlation matrices shown in Appendix~\ref{sec:Supplementary-App}. Four solutions are found, and constraints on the parameter space are shown in Figs.~\ref{fig:gammaOnly}--\ref{fig:gammaDb}. It should be noted that the probability contours shown in the two-dimensional figures do not correspond to $1\sigma,\;2\sigma,\; 3\sigma$, \etc when projected to one dimension.   The preferred solution determines $\gamma$ to be $172^\circ$ with an approximate Gaussian uncertainty of $6^\circ$ and $\delta_{\Bd}^{\D\Kstar}$ to be $296^\circ$ with an approximate Gaussian uncertainty of $8^\circ$. However, the second-preferred solution is consistent with the world-average of direct measurements of $\gamma=\left(66.2^{+3.4}_{-3.6}\right)^\circ$ \cite{HeavyFlavorAveragingGroup:2022wzx}, finding $\gamma$ to be  $62^\circ$ with an approximate Gaussian uncertainty of $8^\circ$  and $\delta_{\Bd}^{\D\Kstar}$ to be $187^\circ$ with an approximate Gaussian uncertainty of $6^\circ$. As can be seen in Fig.~\ref{fig:gammaRb}, all solutions give a consistent determination of $r_{\Bd}^{\D\Kstar}$, with the preferred solution of this analysis finding $r_{\Bd}^{\D\Kstar}=0.235\pm0.017$.

A combined analysis of these results with the measurement of \CP-violating observables from $\Bd\to \D\Kstarz$ with $\D\to \KS h^+h^-$ from Ref.~\cite{LHCb-PAPER-2023-009} is also performed. As the \mbox{$\D\to \KS h^+ h^-$} decay probes $\D$ strong phases other than those close to $0$ and $\pi$,  this breaks the $90^\circ$ degeneracy of the solutions found for $\gamma$ in this analysis and provides additional precision. Projections of the allowed parameter space from the combined analysis are shown in Figs.~\ref{fig:gammaOnly}--\ref{fig:gammaDb}. We estimate the central values and uncertainties of the three primary parameters of interest with the \textsc{Plugin} method of GammaCombo, which implements a frequentist approach. The determinations of the \textsc{Plugin} method including all of the measured $\D$ decay modes in $\Bd\to D\Kstarz$ are $\gamma=\left(63.2^{+6.9}_{-8.1} \right)^\circ$, $r_{\Bd}^{\D\Kstar}=0.233\pm 0.016$, and $\delta_{\Bd}^{\D\Kstar}=\left(192.1^{+6.7}_{-6.1}\right)^\circ$.

In summary, measurements of \CP-violating observables in $\BdorBs\to \D\Kstarz$ decays are presented in this article. These are the most precise measurements to date and provide the most stringent limits to date on $\gamma$ from $\Bd$ decays. The \CP-violating observables in $\Bs$ decays are found to be consistent with no \CP violation. A combined analysis of the \CP-violating observables from \Bd decays finds a solution for $\gamma$ that is consistent with the measurement from $\Bu$ decays reported in the HFLAV average~\cite{HeavyFlavorAveragingGroup:2022wzx}.

\pagebreak
\section*{Acknowledgements}
%
%
\noindent We express our gratitude to our colleagues in the CERN
accelerator departments for the excellent performance of the LHC. We
thank the technical and administrative staff at the LHCb
institutes.
We acknowledge support from CERN and from the national agencies:
CAPES, CNPq, FAPERJ and FINEP (Brazil); 
MOST and NSFC (China); 
CNRS/IN2P3 (France); 
BMBF, DFG and MPG (Germany); 
INFN (Italy); 
NWO (Netherlands); 
MNiSW and NCN (Poland); 
MCID/IFA (Romania); 
MICINN (Spain); 
SNSF and SER (Switzerland); 
NASU (Ukraine); 
STFC (United Kingdom); 
DOE NP and NSF (USA).
We acknowledge the computing resources that are provided by CERN, IN2P3
(France), KIT and DESY (Germany), INFN (Italy), SURF (Netherlands),
PIC (Spain), GridPP (United Kingdom), 
CSCS (Switzerland), IFIN-HH (Romania), CBPF (Brazil),
and Polish WLCG (Poland).
We are indebted to the communities behind the multiple open-source
software packages on which we depend.
Individual groups or members have received support from
ARC and ARDC (Australia);
Key Research Program of Frontier Sciences of CAS, CAS PIFI, CAS CCEPP, 
Fundamental Research Funds for the Central Universities, 
and Sci. \& Tech. Program of Guangzhou (China);
Minciencias (Colombia);
EPLANET, Marie Sk\l{}odowska-Curie Actions, ERC and NextGenerationEU (European Union);
A*MIDEX, ANR, IPhU and Labex P2IO, and R\'{e}gion Auvergne-Rh\^{o}ne-Alpes (France);
AvH Foundation (Germany);
ICSC (Italy); 
GVA, XuntaGal, GENCAT, Inditex, InTalent and Prog.~Atracci\'on Talento, CM (Spain);
SRC (Sweden);
the Leverhulme Trust, the Royal Society
 and UKRI (United Kingdom).

\clearpage
\addcontentsline{toc}{section}{References}
\setboolean{inbibliography}{true}
\bibliographystyle{LHCb}
\bibliography{standard,main,LHCb-PAPER,LHCb-DP,LHCb-CONF}
\clearpage

  \appendix

\section{Correlation matrices}
\label{sec:Supplementary-App}

The statistical correlation matrices of the determined \CP-violating observables are shown in Table~\ref{table:StatCorr}. The systematic correlations are shown in Table~\ref{table:SysCorr}.

\begin{landscape}
\begin{table}[p]
    \caption{Statistical correlation matrix of $B^0$ observables, in percent.}  \label{table:StatCorr}
  \begin{center}
    \begin{tabular}{c|cccccccccccc}
   &$\mathcal{A}_{K\pi}$&$\mathcal{R}_{\pi K}^{+}$&$\mathcal{R}_{\pi K}^{-}$&$\mathcal{A}_{K\pi\pi\pi}$&$\mathcal{R}_{\pi K\pi\pi}^{+}$&$\mathcal{R}_{\pi K\pi\pi}^{-}$&$\mathcal{A}_{CP}^{KK}$&$\mathcal{R}_{CP}^{KK}$&$\mathcal{A}_{CP}^{\pi\pi}$&$\mathcal{R}_{CP}^{\pi\pi}$&$\mathcal{A}_{CP}^{\pi\pi\pi\pi}$&$\mathcal{R}_{CP}^{\pi\pi\pi\pi}$\\\hline
$\mathcal{A}_{K\pi}$&$100$&$9$&$-12$&$0$&$0$&$0$&$0$&$-1$&$0$&$0$&$0$&$0$\\
$\mathcal{R}_{\pi K}^{+}$&&$100$&$6$&$0$&$0$&$0$&$0$&$3$&$0$&$2$&$0$&$0$\\
$\mathcal{R}_{\pi K}^{-}$&&&$100$&$0$&$0$&$0$&$0$&$3$&$0$&$2$&$0$&$0$\\
$\mathcal{A}_{K\pi\pi\pi}$&&&&$100$&$7$&$-5$&$0$&$0$&$0$&$0$&$0$&$-1$\\
$\mathcal{R}_{\pi K\pi\pi}^{+}$&&&&&$100$&$8$&$0$&$0$&$0$&$0$&$0$&$2$\\
$\mathcal{R}_{\pi K\pi\pi}^{-}$&&&&&&$100$&$0$&$0$&$0$&$0$&$0$&$1$\\
$\mathcal{A}_{CP}^{KK}$&&&&&&&$100$&$4$&$0$&$0$&$0$&$0$\\
$\mathcal{R}_{CP}^{KK}$&&&&&&&&$100$&$0$&$5$&$0$&$0$\\
$\mathcal{A}_{CP}^{\pi\pi}$&&&&&&&&&$100$&$4$&$0$&$0$\\
$\mathcal{R}_{CP}^{\pi\pi}$&&&&&&&&&&$100$&$0$&$0$\\
$\mathcal{A}_{CP}^{\pi\pi\pi\pi}$&&&&&&&&&&&$100$&$0$\\
$\mathcal{R}_{CP}^{\pi\pi\pi\pi}$&&&&&&&&&&&&$100$
\end{tabular}
  \end{center}
\end{table}
\end{landscape}

\begin{landscape}
\begin{table}[p]
\centering{}
  \caption{Systematic correlation matrix of $B^0$ observables, in percent.}  \label{table:SysCorr}
\begin{tabular}{c|cccccccccccc}
   &$\mathcal{A}_{K\pi}$&$\mathcal{R}_{\pi K}^{+}$&$\mathcal{R}_{\pi K}^{-}$&$\mathcal{A}_{K\pi\pi\pi}$&$\mathcal{R}_{\pi K\pi\pi}^{+}$&$\mathcal{R}_{\pi K\pi\pi}^{-}$&$\mathcal{A}_{CP}^{KK}$&$\mathcal{R}_{CP}^{KK}$&$\mathcal{A}_{CP}^{\pi\pi}$&$\mathcal{R}_{CP}^{\pi\pi}$&$\mathcal{A}_{CP}^{\pi\pi\pi\pi}$&$\mathcal{R}_{CP}^{\pi\pi\pi\pi}$\\\hline
$\mathcal{A}_{K\pi}$&$100$&$0$&$-5$&$81$&$15$&$-19$&$78$&$-3$&$79$&$-3$&$78$&$-3$\\
$\mathcal{R}_{\pi K}^{+}$&&$100$&$15$&$14$&$7$&$6$&$12$&$5$&$11$&$3$&$5$&$-1$\\
$\mathcal{R}_{\pi K}^{-}$&&&$100$&$-19$&$7$&$4$&$-26$&$14$&$-25$&$4$&$-26$&$-1$\\
$\mathcal{A}_{K\pi\pi\pi}$&&&&$100$&$-1$&$-9$&$77$&$0$&$77$&$-1$&$76$&$-30$\\
$\mathcal{R}_{\pi K\pi\pi}^{+}$&&&&&$100$&$15$&$18$&$0$&$17$&$1$&$19$&$4$\\
$\mathcal{R}_{\pi K\pi\pi}^{-}$&&&&&&$100$&$-24$&$0$&$-24$&$1$&$-23$&$4$\\
$\mathcal{A}_{CP}^{KK}$&&&&&&&$100$&$-3$&$94$&$-2$&$93$&$0$\\
$\mathcal{R}_{CP}^{KK}$&&&&&&&&$100$&$-2$&$15$&$-1$&$-2$\\
$\mathcal{A}_{CP}^{\pi\pi}$&&&&&&&&&$100$&$-3$&$95$&$0$\\
$\mathcal{R}_{CP}^{\pi\pi}$&&&&&&&&&&$100$&$-2$&$0$\\
$\mathcal{A}_{CP}^{\pi\pi\pi\pi}$&&&&&&&&&&&$100$&$-3$\\
$\mathcal{R}_{CP}^{\pi\pi\pi\pi}$&&&&&&&&&&&&$100$

\end{tabular}
\end{table}
\end{landscape}
\clearpage

\centerline
{\large\bf LHCb collaboration}
\begin
{flushleft}
\small
R.~Aaij$^{35}$\lhcborcid{0000-0003-0533-1952},
A.S.W.~Abdelmotteleb$^{54}$\lhcborcid{0000-0001-7905-0542},
C.~Abellan~Beteta$^{48}$,
F.~Abudin{\'e}n$^{54}$\lhcborcid{0000-0002-6737-3528},
T.~Ackernley$^{58}$\lhcborcid{0000-0002-5951-3498},
B.~Adeva$^{44}$\lhcborcid{0000-0001-9756-3712},
M.~Adinolfi$^{52}$\lhcborcid{0000-0002-1326-1264},
P.~Adlarson$^{78}$\lhcborcid{0000-0001-6280-3851},
C.~Agapopoulou$^{46}$\lhcborcid{0000-0002-2368-0147},
C.A.~Aidala$^{79}$\lhcborcid{0000-0001-9540-4988}
% ...

{\footnotesize \it

% ...
$^{35}$Nikhef National Institute for Subatomic Physics, Amsterdam, Netherlands\\
% ...
$^{44}$Instituto Galego de F{\'\i}sica de Altas Enerx{\'\i}as (IGFAE), Universidade de Santiago de Compostela, Santiago de Compostela, Spain\\
% ...
$^{46}$European Organization for Nuclear Research (CERN), Geneva, Switzerland\\
% ...
$^{48}$Physik-Institut, Universit{\"a}t Z{\"u}rich, Z{\"u}rich, Switzerland\\
% ...
$^{52}$H.H. Wills Physics Laboratory, University of Bristol, Bristol, United Kingdom\\
% ...
$^{54}$Department of Physics, University of Warwick, Coventry, United Kingdom\\
% ...
$^{57}$School of Physics and Astronomy, University of Glasgow, Glasgow, United Kingdom\\
$^{58}$Oliver Lodge Laboratory, University of Liverpool, Liverpool, United Kingdom\\
% ...
$^{66}$Syracuse University, Syracuse, NY, United States\\
% ...
$^{78}$Department of Physics and Astronomy, Uppsala University, Uppsala, Sweden, associated to $^{57}$\\
$^{79}$University of Michigan, Ann Arbor, MI, United States, associated to $^{66}$\\
% ...
}
\end{flushleft}

\end{document}